\documentclass[useAMS,usenatbib]{mn2e}
\usepackage{epsfig}
\usepackage{times}
\usepackage{amssymb}
\usepackage{booktabs}
\usepackage{textcomp}
\usepackage{upgreek}

\usepackage{ifpdf}
\ifpdf
\fi

\title[LOFAR 150-MHz observations of SS\,433 and W\,50]{LOFAR 150-MHz observations of SS\,433 and W\,50}
\author[J. W. Broderick et al.]
       {J. W. Broderick,$^{1,2,3}$\thanks{E-mail: broderick@astron.nl} R. P. Fender,$^{1,2}$ J. C. A. Miller-Jones,$^{4}$ S. A. Trushkin,$^{5,6}$ \newauthor A. J. Stewart,$^{1,2}$ G. E. Anderson,$^{1,4}$ T. D. Staley,$^{1,2}$ K. M. Blundell,$^{1}$ M. Pietka,$^{1}$ \newauthor S. Markoff,$^{7}$ A. Rowlinson,$^{3,7}$ J. D. Swinbank,$^{8}$ A. J. van der Horst,$^{9}$ M. E. Bell,$^{10,11}$ \newauthor  R. P. Breton,$^{12}$ D. Carbone,$^{7}$ S. Corbel,$^{13,14}$ J. Eisl\"offel,$^{15}$ H. Falcke,$^{16,3}$ \newauthor J.-M. Grie{\ss}meier,$^{17,14}$ J. W. T. Hessels,$^{3,7}$ V. I. Kondratiev,$^{3,18}$ C. J. Law,$^{19}$ \newauthor G. J. Molenaar,$^{7,20}$  M. Serylak,$^{21,22}$ B. W. Stappers,$^{12}$ J. van Leeuwen,$^{3,7}$ \newauthor R. A. M. J. Wijers,$^{7}$ R. Wijnands,$^{7}$ M. W. Wise$^{3,7}$ and P. Zarka$^{23,14}$                    \\        
       $^{1}$Astrophysics, Department of Physics, University of Oxford, Keble Road, Oxford OX1 3RH, UK\\
       $^{2}$Physics and Astronomy, University of Southampton, Highfield, Southampton SO17 1BJ, UK\\
       $^{3}$ASTRON, the Netherlands Institute for Radio Astronomy, Postbus 2, 7990 AA Dwingeloo, The Netherlands \\  
       $^{4}$International Centre for Radio Astronomy Research, Curtin University, GPO Box U1987, Perth, WA 6845, Australia\\  
       $^{5}$Special Astrophysical Observatory RAS, Karachaevo-Cherkassian Republic, Nizhnij Arkhyz, 369167, Russian Federation \\
       $^{6}$Kazan Federal University, Kazan 420008, Russian Federation \\       
       $^{7}$Anton Pannekoek Institute for Astronomy, University of Amsterdam, Science Park 904, 1098 XH Amsterdam, The Netherlands \\
       $^{8}$Department of Astrophysical Sciences, Princeton University, Princeton, NJ 08544, USA \\  
       $^{9}$Department of Physics, The George Washington University, 725 21st Street NW, Washington, DC 20052, USA \\   
       $^{10}$University of Technology Sydney, 15 Broadway, Ultimo NSW 2007, Australia \\
       $^{11}$ARC Centre of Excellence for All-sky Astrophysics (CAASTRO), The University of Sydney, NSW 2006, Australia \\
       $^{12}$Jodrell Bank Centre for Astrophysics, School of Physics and Astronomy, The University of Manchester, Manchester M13 9PL, UK \\
       $^{13}$Laboratoire AIM (CEA/IRFU - CNRS/INSU - Universit\'e Paris Diderot), CEA DSM/IRFU/SAp, F-91191 Gif-sur-Yvette, France \\
       $^{14}$Station de Radioastronomie de Nan\c{c}ay, Observatoire de Paris, PSL Research University, CNRS, Univ. Orl\'{e}ans, 18330 Nan\c{c}ay, France \\
       $^{15}$Th\"uringer Landessternwarte, Sternwarte 5, D-07778 Tautenburg, Germany \\ 
       $^{16}$Department of Astrophysics/IMAPP, Radboud University Nijmegen, PO Box 9010, 6500 GL Nijmegen, The Netherlands \\
       $^{17}$LPC2E - Universit\'{e} d'Orl\'{e}ans / CNRS, 45071 Orl\'{e}ans cedex 2, France \\
       $^{18}$Astro Space Centre, Lebedev Physical Institute, Russian Academy of Sciences, Profsoyuznaya Str. 84/32, Moscow 117997, Russian Federation \\
       $^{19}$Department of Astronomy and Radio Astronomy Lab, University of California, Berkeley, CA 94720, USA \\
       $^{20}$Department of Physics and Electronics, Rhodes University, PO Box 94, Grahamstown, 6140 South Africa \\ 
       $^{21}$Department of Physics and Astronomy, University of the Western Cape, Private Bag X17, Bellville 7535, South Africa \\
       $^{22}$SKA South Africa, 3rd Floor, The Park, Park Road, Pinelands, 7405, South Africa \\
       $^{23}$LESIA, Observatoire de Paris, CNRS, UPMC, Universit\'e Paris-Diderot, 5 place Jules Janssen, 92195 Meudon, France \\
}

\begin{document}


\pagerange{\pageref{firstpage}--\pageref{lastpage}} \pubyear{2018}

\maketitle
\begin{abstract}
We present LOFAR high-band data over the frequency range $115$--$189$ MHz for the X-ray binary SS\,433, obtained in an observing campaign from 2013 February -- 2014 May. Our results include a deep, wide-field map, allowing a detailed view of the surrounding supernova remnant W\,50 at low radio frequencies, as well as a light curve for SS\,433 determined from shorter monitoring runs. The complex morphology of W\,50 is in excellent agreement with previously published higher-frequency maps; we find additional evidence for a spectral turnover in the eastern wing, potentially due to foreground free-free absorption. Furthermore, SS\,433 is tentatively variable at $150$ MHz, with both a debiased modulation index of $11$ per cent and a $\chi^2$ probability of a flat light curve of $8.2 \times 10^{-3}$. By comparing the LOFAR flux densities with contemporaneous observations carried out at $4800$ MHz with the RATAN-600 telescope, we suggest that an observed $\sim$$0.5$--$1$ Jy rise in the $150$-MHz flux density may correspond to sustained flaring activity over a period of approximately six months at $4800$ MHz. However, the increase is too large to be explained with a standard synchrotron bubble model. We also detect a wealth of structure along the nearby Galactic plane, including the most complete detection to date of the radio shell of the candidate supernova remnant G\,38.7$-$1.4. This further demonstrates the potential of supernova remnant studies with the current generation of low-frequency radio telescopes.    
\end{abstract}

\label{firstpage}

\section{Introduction}\label{introduction}

The W\,50 nebula \citep[][]{westerhout58} is a large, non-thermal, Galactic radio source, at the centre of which is the X-ray binary system SS\,433 \citep[][]{stephenson77}, a known source of relativistic jets \citep[][]{fabian79,margon79a,margon79b,liebert79,milgrom79,abell79}. These jets, with mean velocity $0.26$$c$, precess about an axis with a period of $162.4$ d, where the half-opening angle of the precession cone is $21^{\circ}$ and the precession axis is inclined at $78^{\circ}$ to the line of sight (\citealt[][]{eikenberry01}; also see \citealt[][]{margon84}). This creates a corkscrew-shaped trace on the sky as the jet components move ballistically outwards \citep[][]{hjellming81a,hjellming81b}. 

At radio frequencies, W\,50 takes the form of a circular shell of diameter $58$ arcmin with lateral extensions (the `wings' or `ears') of $47$ arcmin and $25$ arcmin to the east and west, respectively \citep[e.g. $327.5$- and $1465$-MHz maps in][henceforth D98]{dubner98}. Originally classified as a supernova remnant (SNR) by \citet[][]{holden69}, it was suggested with the discovery of the jets in SS\,433 that, given the remarkable alignment between the jet axis and the major axis of W\,50, they are responsible for the elongated ($\sim$ $2\degr \times 1\degr$) shape of the nebula \citep[][]{begelman80}. Hydrodynamical simulations by \citet*[][]{goodall11a} suggested that W\,50 is about $20\,000$ years old, and that several distinct episodes of jet activity, with different precession characteristics to the current jets, can explain the morphology of W\,50; this is consistent with the kinematical investigation by \citet*[][]{goodall11b}. The asymmetry in the sizes of the ears is usually ascribed to the known density gradient associated with the Galactic plane; SS~433 is located at Galactic coordinates $(l,b) = (39\fdg7, -2\fdg2)$, and the major axis of W\,50 is oriented at $20$$^{\circ}$ from the normal to the plane (e.g. \citealt*[][]{downes81a}; \citealt[][]{goodall11a}). 

In observations at frequencies $\gtrsim 1$ GHz, flaring activity from SS\,433 has been found to occur on time-scales $\lesssim 100$--$200$ days \citep*[e.g.][]{seaquist82,fiedler87,vermeulen93,trushkin03,pal06,trushkin12,trushkin14,trushkin16}. During flaring, flux densities of $\sim$$0.5$--$4$ Jy are measured, on average a factor of a few above the quiescent baseline value. Flaring has also been detected at $408$ and $843$ MHz \citep[e.g.][]{bonsignori86,vermeulen93}; at $408$ MHz, the source is typically about $1$ Jy brighter than during quiescence \citep[$3.4$ Jy versus $2.3$ Jy;][]{bonsignori86}.    

At low radio frequencies ($< 300$ MHz), however, SS\,433, and indeed W\,50 as well, have been relatively poorly characterized. A brief summary of previous observations is as follows. Firstly, using $80$- and $160$-MHz data from the Culgoora Circular Array (CCA), as well as near-simultaneous higher-frequency measurements, \citet[][]{seaquist80} suggested that the quiescent radio spectrum of SS\,433 turns over at approximately $300$ MHz. From two epochs, spread three weeks apart, the average $160$-MHz flux density was $1.9$ Jy, and the average $5\sigma$ upper limit at $80$ MHz was $1.6$ Jy. Secondly, \citet[][]{pandey07} monitored SS\,433 at $235$ MHz with the Giant Metrewave Radio Telescope (GMRT) over the period 2002 July -- 2005 January (observational cadence range $2$--$270$ days). The average $235$-MHz flux density was $2.71$ Jy and the modulation index $21$ per cent. Using simultaneous $610$-MHz GMRT observations, the two-point spectral index\footnote{In this paper, we use the convention $S_{\nu} \propto \nu^{\alpha}$, where $S_{\nu}$ is the flux density at frequency $\nu$, and $\alpha$ is the spectral index.}  ${\alpha}^{610}_{235}$ varied between $-1.05$ and $-0.69$, where the uncertainty per individual measurement is much less than this range. This is possibly due to optical depth variations at different stages of flaring \citep[discussion in][]{pandey07}. Lastly, \citet[][]{millerjones07} detected both SS\,433 and W\,50 with the Very Large Array (VLA) at $74$ MHz; the root-mean-square (rms) noise level in their map was $192$ mJy beam$^{-1}$, with an angular resolution of about $100$ arcsec. The flux density of SS\,433 was $1.9 \pm 0.2$ Jy (J. Miller-Jones, priv. comm.). Using these data, as well as previously published observations, including $83$- and $102$-MHz measurements from the Pushchino Radio Astronomy Observatory \citep*[][]{kovalenko94}, \citet[][]{millerjones07} demonstrated that the integrated radio spectrum of W\,50 can be modelled with a single power law between $74$ and $4750$ MHz, where $\alpha=-0.51 \pm 0.02$ \citep[also see D98 and][]{gao11}.

With the advent of a new generation of novel, low-frequency radio telescopes, including the Low-Frequency Array \citep[LOFAR;][]{vanhaarlem13}, the Murchison Widefield Array \citep[MWA;][]{tingay13} and the Long Wavelength Array \citep[LWA;][]{ellingson13}, there is now an excellent opportunity to embark on a much more thorough exploration of the low-frequency synchrotron radio emission from jet-producing sources such as SS\,433. Topics of particular interest include constraining models for intrinsic variability over wide frequency ranges (including frequency-dependent time lags), investigating absorption processes that result in a spectral turnover, as well as the presence of a low-frequency cutoff in the synchrotron spectrum. In this paper, we present the results of a LOFAR high-band observing campaign of SS433 and W50, carried out over the frequency range $115$--$189$ MHz during the period 2013 February -- 2014 May. In Section~\ref{observations}, we describe the observations, and how the data were calibrated and imaged. A deep continuum map is presented in Section~\ref{deep map}, while in Section~\ref{spectral index discussion} we combine our data with previous observations at lower and higher frequencies to investigate the radio spectra of the components of W\,50. Our low-frequency SS\,433 monitoring observations are analysed in Section~\ref{SS433 variability}, including a comparison with contemporaneous data at $4800$ MHz from the RATAN-600 telescope. Then, in Section~\ref{SNR candidate}, we discuss the detection of a nearby candidate SNR, G\,38.7$-$1.4, in our deep image. Finally, we conclude in Section~\ref{conclusions}.      

\section[]{LOFAR observations and data reduction}\label{observations}

A detailed description of LOFAR can be found in \citet[][]{vanhaarlem13}. In this section, we provide a summary of our observations, carried out with the high-band antennas (HBA; full operating frequency range $110$--$240$ MHz). An observing log is given in Table~\ref{table:observations}.    

\begin{table*}
\begin{minipage}{140mm}
 \centering
  \caption{LOFAR high-band observations of SS\,433 and W\,50. MJD is the Modified Julian Date at the halfway point of each observation. The prefix `M' in column 1 refers to the monitoring observations. See Section~\ref{SS433 variability} for a description of how the SS\,433 flux densities were obtained; uncertainties are $\pm 1 \sigma$. The on-source integration time is also given for each run, as well as the LOFAR observation IDs.}
  \begin{tabular}{llccrcr}
  \hline
   \multicolumn{1}{c}{Run}  &  \multicolumn{1}{c}{Date} & \multicolumn{1}{c}{MJD} & \multicolumn{1}{c}{$S_{\rm{SS}433}$} & $t_{\rm int}$ & \multicolumn{1}{c}{No. stations} & \multicolumn{1}{c}{IDs} \\    
& & \multicolumn{1}{c}{} & \multicolumn{1}{c}{(Jy)} & \multicolumn{1}{c}{(min)} & \multicolumn{1}{c}{(core, remote, intl.)} & \multicolumn{1}{c}{(range)}  \\
 \hline
 Deep & 2013 February 18 & $56341.38$ & $2.08 \pm 0.21$ & $176$ & $24$, $13$, $0$ & L$95041$--L$95072$ \\
 M1 & 2013 March 20 & $56371.28$ & $2.09 \pm 0.21$ & $22$ & $24$, $13$, $0$ & L$106054$--L$106057$ \\
 M2 & 2013 April 25 & $56407.22$ & $1.65 \pm 0.17$ & $22$ & $16$, $13$, $8$\rlap{$^{a}$}  & L$128512$--L$128515$ \\
 M3 & 2013 May 23 & $56435.12$ & $1.69 \pm 0.17$ & $22$ & $22$, $13$, $0$ & L$136070$--L$136073$ \\
 M4 & 2013 July 25/26 & $56499.00$ & $1.49 \pm 0.15$ & $22$ & $19$, $13$, $0$ & L$166060$--L$166063$  \\
 M5 & 2013 August 22  & $56526.87$ & $1.60 \pm 0.16$ & $22$ & $24$, $14$, $0$ & L$169277$--L$169280$  \\
 M6 & 2013 September 20 & $56555.86$ & $1.74 \pm 0.18$ & $22$ & $24$, $14$, $0$ & L$175340$--L$175343$  \\
 M7 & 2013 September 28 & $56563.83$ & $1.64 \pm 0.17$ & $22$ & $24$, $14$, $0$ & L$178910$--L$178913$ \\
 M8 & 2013 November 16 &  $56612.63$ & $2.14 \pm 0.21$ & $19$\rlap{$^{b}$}  & $24$, $14$, $0$  & L$188650$--L$188651$ \\ 
 M9 & 2013 December 11 & $56637.55$ &  $2.13 \pm 0.21$ & $26$ & $24$, $14$, $0$ & L$193989$--L$193992$ \\ 
 M10 & 2014 February 11 & $56699.39$ & $2.62 \pm 0.26$ & $26$ & $24$, $11$, $0$  & L$204224$--L$204227$ \\
 M11 & 2014 March 12 & $56728.30$ &  $2.06 \pm 0.21$ & $26$  & $24$, $14$, $0$  & L$207137$--L$207140$ \\
 M12 & 2014 April 24 & $56771.25$ &  $1.89 \pm 0.19$ & $26$ & $24$, $14$, $0$  & L$215295$--L$215298$  \\ 
 M13 & 2014 May 10 & $56787.10$ &  $1.89 \pm 0.19$ & $26$  & $24$, $14$, $0$  & L$228277$--L$228280$  \\
\hline
\multicolumn{7}{l}{Note: additional monitoring observations on 2013 September 16 and 2014 January 8 were unusable due to various}\\  
\multicolumn{7}{l}{observational and data quality issues.}\\  
\multicolumn{7}{l}{$^{a}$The observation on 2013 April 25 included eight international stations in place of an equivalent number of core}\\ 
\multicolumn{7}{l}{stations, potentially allowing very-high-resolution images to be generated, but we only consider the Dutch stations}\\  
\multicolumn{7}{l}{in this paper.}\\  
\multicolumn{7}{l}{$^{b}$This snapshot was obtained as part of a second deep observation, but of lower data quality. The `HBA Dual Inner'}\\ 
\multicolumn{7}{l}{configuration was used, in which the primary beam FWHM of the remote stations is matched to the core}\\ 
\multicolumn{7}{l}{sub-stations by using the inner $24$ tiles of the remote stations only.}\\ 
\end{tabular}
\label{table:observations}
\end{minipage}
\end{table*}

\subsection{Deep continuum observation and monitoring runs} 

Our first HBA observation of SS\,433 and W\,50 was a deep $4$-h run carried out with the `HBA Dual' mode, centred on transit (elevation $42\degr$). The frequency range was $115$--$163$ MHz, covered by a total of $244$ sub-bands, each with a bandwidth of $195.3$ kHz. We used the Dutch stations only: $24$ stations in the core, each with two closely-spaced sub-stations, and $13$ remote stations. The projected baselines range from about $29$ m to $79$ km. The primary beam full width at half maximum (FWHM) at $150$ MHz is $3\fdg80$ for the core sub-stations and $2\fdg85$ for the remote stations.   

The observation was carried out in blocks of $15$ min: $2$ min on the primary flux density calibrator, 3C\,380, followed by $11$ min on the target. Although LOFAR is a software telescope, rapid electronic beam switching was not possible when our study took place: the remaining $2 \times 1$ min in the block was needed to switch between the calibrator and target (and vice versa). By the end of the run, $16$ snapshots of SS\,433 and W\,50 had been obtained, corresponding to a total integration time of $176$ min. 

The primary goal of the shorter runs was to monitor the flux density of SS\,433. The setup was very similar, except that the frequency coverage was extended to $115$--$189$ MHz, spread over $380$ sub-bands. In general, $2 \times 11$-min or $2 \times 13$-min snapshots, near transit, were obtained per run, along with $2 \times 2$-min scans of 3C\,380. However, the data from 2013 November 16 were obtained with a slightly different setup: one $19$-min snapshot, approximately centred on transit, preceded by a $1.5$-min observation of 3C\,380.

\subsection{Pre-processing, calibration and imaging}\label{calibration and imaging} 

For all observations, data were recorded with a time-step of $2$ s and $64$ channels per sub-band. Pre-processing was carried out using standard methods \citep[e.g.][]{vanhaarlem13}. Firstly, radio-frequency interference (RFI) was removed using {\sc aoflagger} \citep*[][]{offringa10,offringa12}, and the two edge channels at both ends of each sub-band were completely flagged, reducing the bandwidth per sub-band to an effective value of $183.1$ kHz. Secondly, because the very bright `A-team' source Cygnus A is only $17\fdg$8 from 3C\,380, the `demixing' algorithm \citep*[][]{vandertol07} was used to subtract the response of the former from the visibilities of the latter. Lastly, for practical reasons concerning data volume, and the computing time required for calibration and imaging, the data were averaged in time and frequency: to $10$ s per time-step and either $4$ frequency channels per sub-band (deep run) or $1$ frequency channel per sub-band (monitoring runs). For the baseline ranges considered in this paper, the effects of bandwidth and time-average smearing \citep[e.g.][]{bridle99} in the averaged data are well within the calibration uncertainty (Section~\ref{in band flux scale}), even at the edges of the field. 

Calibration and imaging were also carried out using standard practices. Each 3C\,380 sub-band was calibrated using a model of the source defined in \citet{scaife12}; the absolute flux density scale is that of \citet*[][]{roger73}. The gain amplitudes were transferred to the target sub-bands in the same observing block. We then performed phase-only calibration on the target field using data from the global sky model developed by \citet[][]{scheers11}, within a radius of $10$ deg from the position of SS\,433. The basis for our model of the field was the $74$-MHz VLA Low-Frequency Sky Survey \citep[VLSS(r);][]{cohen07,lane14}, with spectral index information being obtained by cross-correlating the relevant VLSS catalogue entries with the $1.4$-GHz NRAO VLA Sky Survey \citep[NVSS;][]{condon98}. To increase the signal-to-noise and hence the accuracy of the phase-only calibration step, solutions were computed for at most $20$ sub-bands at a time ($\Delta\nu \approx 4$ MHz). Furthermore, we restricted the projected baseline range over which the phase solutions were determined to $250$--$5000$ $\lambda$ (i.e. $500$ m -- $10$ km at $150$ MHz), so as to avoid the shortest and longest baselines where our sky model will generally be an inadequate representation of the true sky brightness distribution. 

Prior to imaging each run, the sub-bands were combined into six, approximately evenly split bands, and the data were visually inspected; any further errant data points were manually flagged. A primary-beam-corrected image was then generated for each band with {\sc awimager} \citep[][]{tasse13}; we used a robust weighting parameter \citep[][]{briggs95} of $0$. At the time of analysis, the {\sc awimager} did not fully account for variations of flux density with frequency in the deconvolution process \citep[e.g.][]{sault94}; by making multiple images across the full bandwidth, we kept the fractional bandwidth of the data being imaged to $<10$ per cent in all cases, which helps to minimize the possible reduction in image quality due to effects associated with this issue.  

For the deep run, a maximum projected baseline of $4000$ $\lambda$ was used when imaging; for the monitoring runs, the baseline range was $100$--$3000$ $\lambda$. In the latter case, the lower cutoff reduced the contribution from the diffuse emission from W\,50, while both higher cutoffs gave the most reliable images for the given $uv$ coverage and relatively simple calibration procedure. A multi-scale {\sc clean} option had not yet been implemented in the {\sc awimager}; it will be possible in the future to more accurately image the significant amount of extended emission in this field.  

To make a final map for each run, the six separate images were first convolved to a common angular resolution using the task {\sc convol} in {\sc miriad} \citep*[][]{sault95}. We then stacked and averaged these resulting maps together with the {\sc miriad} tasks {\sc imcomb} and {\sc avmaths}. Equal weights were used; the effective frequency is about $140$ MHz for the deep run and $150$ MHz for the monitoring runs.  

Lastly, after a first pass at imaging, we included an additional round of phase-only self-calibration with a model generated from the LOFAR image. The source finder {\sc pybdsf} \citep[][]{mohan15} was used to generate the model; it was restricted to sources sufficiently compact such that they could be well described with either a single Gaussian or a moderate number ($\leq 6$) of Gaussians. While we subsequently used these self-calibration solutions (determined over the same projected baseline range as before), we found that they did not make a significant difference to the image properties.

\subsection{In-band flux density accuracy}\label{in band flux scale}

At the time of analysis, an issue existed with the standard LOFAR HBA beam model such that the in-band spectral index could be artificially steep. The magnitude of the effect is dependent on the elevation of the target in a given observation, and the distance between the primary calibrator and the target. Unfortunately, this currently precludes an investigation of how the radio properties of SS\,433 and W\,50 potentially vary within the LOFAR high band. However, to ensure that the average HBA flux density scale is as accurate as possible, approximate corrections can be made based on a preliminary analysis of full simulations of the electromagnetic properties of the dipoles in the array, including the effects of mutual coupling and grating lobe attenuation (G. Heald, priv. comm.).

For the deep observation, such corrections were applied to each of the six images before they were convolved to a common resolution and averaged together; we divided by factors ranging from $1.6$--$0.8$ between $115$--$163$ MHz. To first order, this process removes an artificially steep in-band spectral index of approximately $-2.5$ in this case. Unfortunately, the simulations are currently incomplete above $163$ MHz, meaning that we are not able to fully assess the effect on the monitoring runs. However, based on an extrapolation of the available data between $115$--$163$ MHz, we estimate that the average correction across a bandwidth of $115$--$189$ MHz should, to first order, fall within the general calibration uncertainty: for all runs, based on the flux stability of sources in the target field, along with 3C\,380 itself, we estimate that the internal flux density calibration uncertainty is $\sim$$10$ per cent. Hence, we have made no flux density scale adjustments to the images from the monitoring runs.

\subsection{Absolute flux density calibration}\label{absolute flux calibration}

We also cross-checked the band-averaged LOFAR flux densities of the SNRs 3C\,396 and 3C\,397 (also known as G\,039.2$-$00.3 and G\,041.1$-$00.3, respectively; e.g. \citealt[][]{green14}), as well as 4C\,$+$06.66, with flux densities measured at $160$ MHz with the CCA \citep[][]{slee95}. Using the information presented in \citet[][]{slee95} and \citet[][]{scaife12}, both our LOFAR data, and the data from \citet[][]{slee95}, are tied to absolute flux density scales that agree to within $\sim$$3$ per cent. We measured the LOFAR flux densities from our deep image, which has the more complete $uv$ and baseline coverage, and, additionally, we interpolated the $160$-MHz CCA flux densities to $140$ MHz using the $80$-MHz CCA flux densities and two-point spectral indices $\alpha^{160}_{80}$ that were also reported in \citet[][]{slee95}.  

We find that our LOFAR flux densities for 3C\,396 and 4C\,$+$06.66 agree with the interpolated $140$-MHz flux densities from \citet[][]{slee95} to within approximately $\pm 2$ per cent, giving us confidence in the absolute flux density calibration. However, the LOFAR measurement for 3C\,397 is about $25$ per cent lower. One possibility to explain at least some of this discrepancy is that the \citet[][]{slee95} measurement at $160$ MHz has recorded more diffuse emission from the source, and/or there is significant confusion in this case from the Galactic plane, owing to the lower angular resolution, which is roughly twice as coarse as in our study. 

To investigate further whether this finding for 3C\,397 is a potential outlier, we conducted a similar comparison for 3C\,398 (G\,043.3$-$00.2), which we also detect, albeit well into a sidelobe of the primary beam; this source is $4\fdg1$ to the north of SS\,433. Despite the fact that a flux density measured so far from the phase centre may inherently be unreliable, interestingly the agreement with the adjusted measurement from \citet[][]{slee95} is well within the $\pm 2$ per cent margin described above for 3C\,396 and 4C\,$+$06.66.   

Given the above results for 3C\,396 and 4C\,$+$06.66 (and 3C\,398), we conclude that our data have been correctly tied to the \citet[][]{roger73} absolute flux density scale, and that the main calibration uncertainty is an internal one, as outlined above in Section~\ref{in band flux scale}. Moreover, we will show throughout this paper that there is a good agreement between our LOFAR flux densities and previously published flux density measurements. Of our key conclusions (Section~\ref{conclusions}), further evidence for spectral curvature in the eastern wing of W\,50 (Section~\ref{spectral index discussion}), as well as the corresponding fit parameters, may be affected if the calibration uncertainty is larger than we have calculated.  

Generally speaking, it is important to note that absolute flux density calibration is well known to be challenging at low frequencies with aperture arrays such as LOFAR, particularly near and along the Galactic plane. Due to the excellent sensitivity to diffuse extended emission, future planned releases of Galactic plane catalogues from the LOFAR Multifrequency Snapshot Sky Survey \citep[MSSS;][]{heald15}, and the GaLactic and Extragalactic All-sky Murchison Widefield Array survey \citep[GLEAM;][]{hurleywalker17}, will be highly valuable comparison datasets.

\section{A LOFAR map of SS\,433 and W\,50}\label{deep map}

\begin{figure*}
\begin{minipage}{180mm}
\centering
\includegraphics[width=18.5cm]{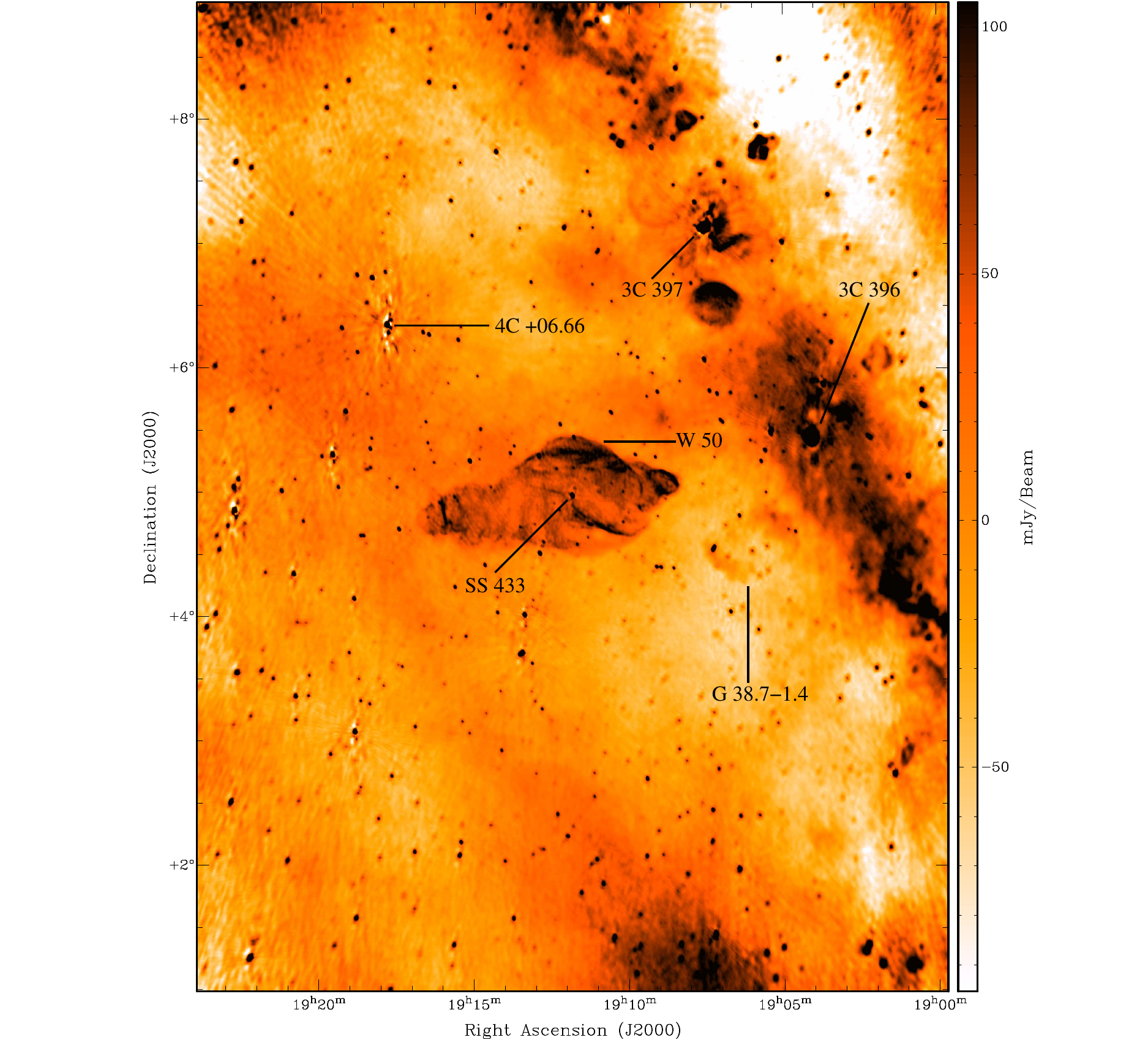}
\caption{Deep LOFAR $140$-MHz high-band continuum map, with dimensions $6.0$ deg $\times$ $7.9$ deg, centred on SS\,433 and W\,50. The angular resolution is $78$ arcsec $\times$ $55$ arcsec (BPA $20^{\circ}$), and the pixels are $10$ arcsec $\times$ $10$ arcsec in size. We have imaged beyond the half-power points of the core and remote station primary beams, which are $4\fdg1$ and $3\fdg0$ FWHM, respectively, for an average frequency of $140$ MHz. The colour bar range does not include all pixel values (maximum $10.4$ Jy beam$^{-1}$): the contrast level has been chosen to show the significant amount of extended emission in this field, and many sources, including SS\,433, are saturated using this scheme. SS\,433 and W\,50 are labelled, as well as several other sources that are discussed in the text.}
\label{fig:deepmap}
\end{minipage}
\end{figure*}

\begin{figure*}
\begin{minipage}{180mm}
\centering
\includegraphics[width=17cm]{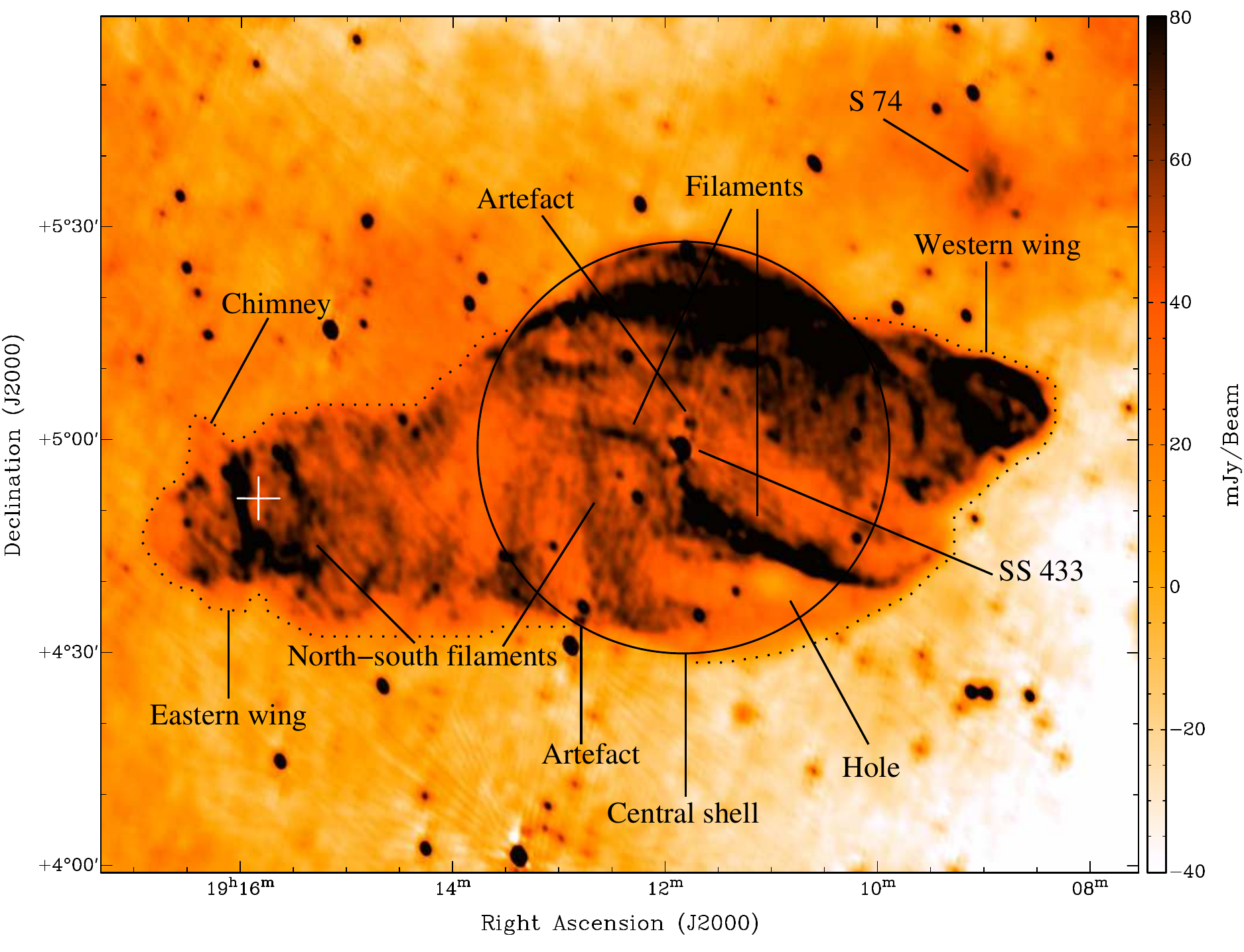}
\caption{The map from Figure~\ref{fig:deepmap}, zoomed in on the region containing SS\,433 and W\,50. We define the central shell as in D98: a circle, radius $29$ arcmin, centred on SS\,433. Note, however, that the true centre of W\,50 has been determined to be approximately $5$ arcmin to the east-south-east of SS\,433, along the jet axis (\citealt[][]{watson83}; \citealt*[][]{lockman07}; \citealt[][]{goodall11a}). The eastern and western wings are labelled; the dotted lines approximately enclose the regions from which we measured the flux densities (discussion in Section~\ref{W50 fluxes}). We also indicate the positions of a selection of filaments, the spur (or `chimney'), two artefacts, and the H\,{\sc ii} region S\,74 (discussion in Sections~\ref{W50 morphology} and \ref{nearby features}). The white cross refers to the $327.5$-MHz position of a variable point source (Section~\ref{variability nearby sources}). As in Figure~\ref{fig:deepmap}, the contrast level is chosen to show the significant amount of extended emission from W\,50.}
\label{fig:deepmap2}
\end{minipage}
\end{figure*}

Our deep continuum map is presented in Figure~\ref{fig:deepmap}, with a zoomed-in view of SS\,433 and W\,50 shown in Figure~\ref{fig:deepmap2}. This is the most detailed image of W\,50 made thus far at frequencies below $300$ MHz. The angular resolution is $78$ arcsec $\times$ $55$ arcsec (beam position angle BPA $= 20^{\circ}$; throughout this paper we use the convention that the BPA is measured north through east), comparable to that of the $327.5$- and $1465$-MHz VLA images from D98; in particular, the resolution of the latter map from D98 is $56$ arcsec $\times$ $54$ arcsec (BPA $-61\fdg4$). Indeed, at a similar angular resolution, there is a clear consistency between the LOFAR and D98 maps: we also observe the three main components of the nebula (i.e. the central shell, as well as the eastern and western wings) and the complex structure contained within. A similar correspondence can be seen when comparing the LOFAR data with the recent study by \citet[][]{farnes17}, who conducted a full-Stokes broadband analysis of SS\,433 and W\,50, centred at about $2.3$ GHz, with the Australia Telescope Compact Array (ATCA); the angular resolution of their map is $2.5$ arcmin $\times$ $1.9$ arcmin (BPA $75\fdg9$). Within the wide field of view, we have also detected a wealth of structure near and along the Galactic plane; we return to this topic in Section~\ref{SNR candidate}.  

In the vicinity of W\,50, the rms noise level is $\sim$$4$--$6$ mJy beam$^{-1}$. Source and sidelobe confusion so near the Galactic plane, together with the presence of undersampled diffuse emission, are responsible for the noise level being much higher than the theoretical value of $\sim$$0.08$ mJy beam$^{-1}$. Moreover, there are clear variations off-source (henceforth referred to as the `background'), including `negative bowls', due to the undersampling of extended emission.

\subsection{Flux densities}\label{W50 fluxes}  
   
A Gaussian fit (using {\sc imfit} in {\sc miriad}) to SS\,433 yields a peak flux density of $2.01 \pm 0.20$ Jy beam$^{-1}$ and an integrated flux density of $2.25 \pm 0.23$ Jy. Note that the $10$ per cent calibration uncertainty is the dominant error term. There is very marginal evidence of source extension (at about the $1.6\sigma$ level along both the major and minor axes of the fitted Gaussian), but the deconvolved position angle, approximately $30^{\circ}$, is not consistent with the jet axis and precession cone. Intriguingly, however, this position angle is consistent with the general direction of the low-surface-brightness `ruff' emission observed at GHz frequencies \citep[e.g.][]{blundell01}, albeit on much smaller angular scales (tens of mas) than in our map.        

Assuming that W\,50 has a central shell with radius $29$ arcmin (D98), we calculated the integrated flux densities of the various components of the nebula by summing the pixels in the respective regions (indicated in Figure~\ref{fig:deepmap2}) and making corrections for the non-zero background levels. Each background level correction value was determined by measuring the mean flux density in a series of boxes around the component and then taking the average of this set of means; moreover, the standard deviation of the set was used as the uncertainty for the correction. The mean offsets that were subtracted (per pixel) range from $-15.5$ to $6$ mJy beam$^{-1}$. Twisted-plane fits were also investigated, but the results were deemed to be less reliable. Henceforth in this section, as well as Sections~\ref{W50 morphology} and \ref{nearby features}, all quoted integrated flux densities and surface-brightness values are background corrected, unless stated otherwise. 

The integrated flux densities for the eastern wing, central shell and western wing are $48 \pm 13$, $121 \pm 25$ and $38 \pm 8$ Jy, respectively. The flux density errors were calculated by combining the $10$ per cent calibration uncertainty and the uncertainty in the background level in quadrature; there is also a much smaller statistical uncertainty from the noise level. The $2.25$-Jy contribution from SS\,433 was subtracted from the central shell, but we do not make any corrections for possibly unassociated compact sources in the three regions; their effects are essentially negligible in comparison to the other flux density uncertainties. 

Summing the integrated flux densities for the three components stated in the previous paragraph, and taking into account the relative contributions of systematic and statistical uncertainties when propagating the errors, our total integrated flux density for W\,50 at $140$ MHz is $207 \pm 33$ Jy. This value is consistent with the previously calculated integrated radio spectrum (see discussion in Section~\ref{introduction}), giving us further confidence in our calibration and background subtraction procedures. However, depending on the relative contributions of the negative bowls and diffuse Galactic emission in Figures~\ref{fig:deepmap} and \ref{fig:deepmap2}, one possibility is that the mean background corrections should be more negative in value (i.e. the flux densities of the three components of W\,50 should be higher), and that the respective uncertainties are overestimated. As will become apparent in Section~\ref{integrated flux densities spectra}, this would move the flux densities of the eastern wing and central shell upwards towards the extrapolated single power-law radio spectra determined by D98 at higher frequencies. In the former case, this would reduce the evidence for spectral curvature at low frequencies. However, it seems very unlikely that an incorrect background correction alone could explain the deviation from a single power-law radio spectrum for the eastern wing, as there is insufficient evidence in Figures~\ref{fig:deepmap} and \ref{fig:deepmap2} for the significant negative bowl (mean $\approx -46$ mJy beam$^{-1}$) that would be needed in the vicinity of this component. Furthermore, if the background correction for the western wing were to be a larger negative value, then the corrected $140$-MHz flux density would quickly become overestimated.

Various second-order effects may result from the use of the \citet[][]{clark80} {\sc clean} algorithm in the {\sc awimager} on a target such as W\,50, including residual deconvolution sidelobes and artefacts, as well as possible {\sc clean} bias \citep[e.g.][]{condon98}. We plan to re-image our data with multi-scale {\sc clean} and maximum entropy algorithms once they are implemented in the {\sc awimager}; an alternative would be to use sparse image reconstruction methods \citep[e.g.][]{garsden15,dabbech15,girard15,pratley18}. Other techniques, such as the one outlined in \citet[][]{braun85}, could also be considered as part of future work.

\subsection{Low-frequency morphology of W\,50}\label{W50 morphology}

\begin{figure}
\centering
\includegraphics[width=8.5cm]{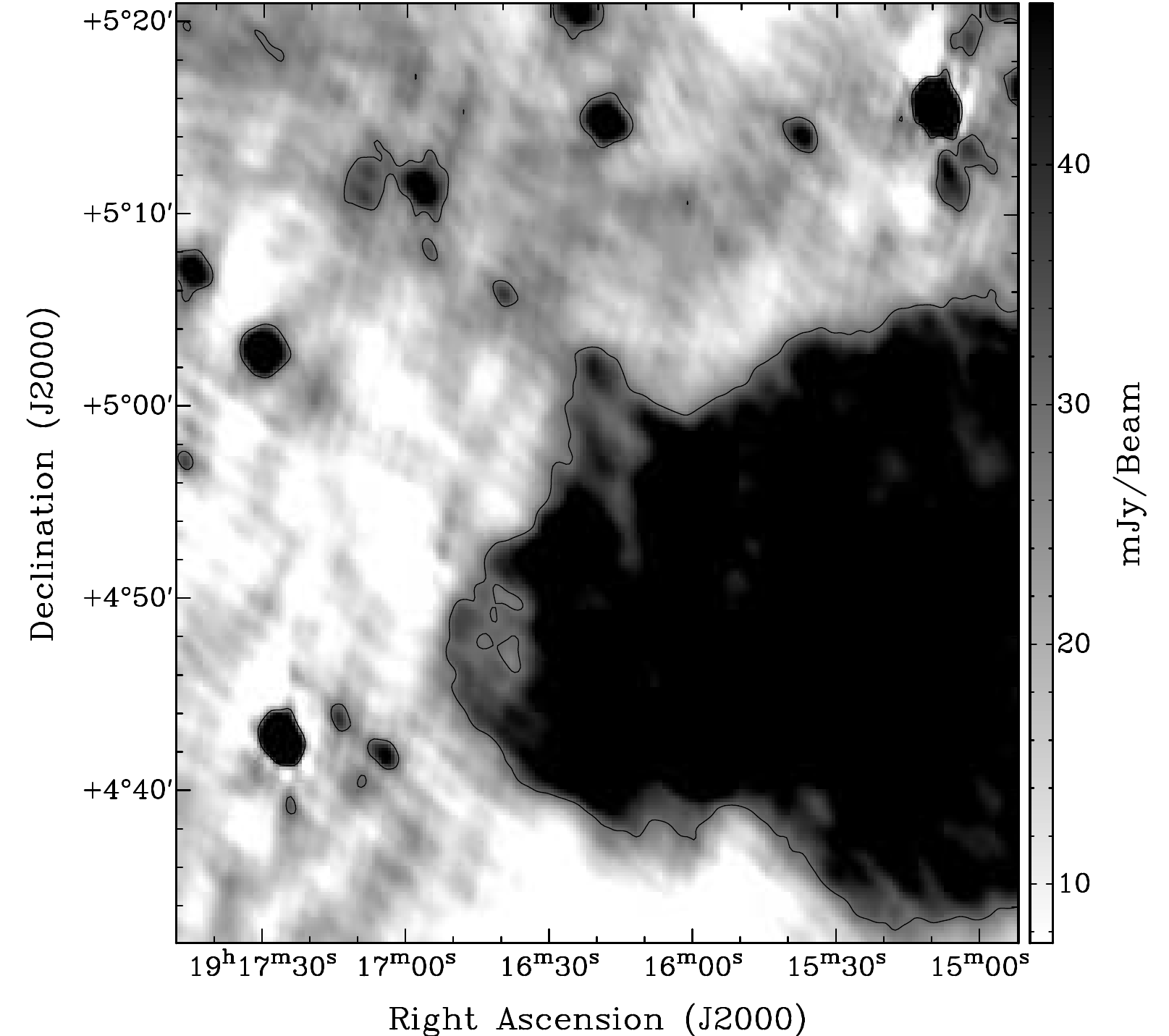}
\caption{The spur, or `chimney', at $140$ MHz. To outline its shape, we have overlaid a radio contour at the $30$ mJy beam$^{-1}$ level (not corrected for the non-zero background).}
\label{fig:chimney}
\end{figure}

The arcs and filaments observed at higher frequencies are also prominent in our map. These include the thin filament at a position angle of $80^{\circ}$ (labelled in Figure~\ref{fig:deepmap2}), discussed in D98 as being positionally associated with the precession cone of SS\,433, with a $140$-MHz surface brightness ranging from about $50$--$90$ mJy beam$^{-1}$. In addition, we see the set of north-south filaments which are suggestive of large-scale traces of the precessing jets; the brighter of these are also labelled in Figure~\ref{fig:deepmap2}, with hints of similar fainter features in between. The peak flux densities of the eastern and western wings are approximately $150$ and $170$ mJy beam$^{-1}$, respectively; they occur in the brightest north-south filament and near the edge-brightened western boundary, respectively.  

We detect the `chimney' (Figure~\ref{fig:chimney}), previously observed at $1465$ MHz by D98 and also seen by \citet[][]{farnes17}. This is a spur-like feature to the north of the bright filament in the eastern wing, near the position RA $19^{\rm h}$\,$16^{\rm m}$\,$18^{\rm s}$, Dec. $+04^{\circ}$\,$59\arcmin\,30\arcsec$ (J2000). D98 suggested that it may be analogous in morphology to a spur, approximately an order of magnitude smaller, at the northern edge of the Crab Nebula \citep[e.g.][]{fesen86,rudie08,black15}. The approximate dimensions at $140$ MHz are very similar to those at $1465$ MHz: $\sim$$6$ arcmin in length and $\sim$$7$ arcmin in width at the base. This corresponds to a projected linear size of about $10$ pc $\times$ $11$ pc.\footnote{In this paper, we assume that the distance to SS\,433 and W\,50 is $5.5$ kpc \citep[][]{blundell04,lockman07}. At this distance, an angular size of $1$ arcmin corresponds to a projected linear size of $1.6$ pc.} The average surface brightness at $140$ MHz is $\sim$$15$--$20$ mJy beam$^{-1}$. Using the $1465$-MHz image from D98, convolved to the resolution of our data (see Section~\ref{spectral index discussion map} for further details, where we construct a two-point spectral index $\alpha^{1465}_{140}$ map of W\,50), we estimate a two-point spectral index $\alpha^{1465}_{140} \sim -0.5$, with an error of about $30$--$40$ per cent due to the faintness of the emission and the uncertainties in estimating the background level at both frequencies. 

\citet[][]{farnes17} identified a potential second chimney, $20$ arcmin to the west of the chimney described above. However, it is not visible in Figure~\ref{fig:chimney}. Our non-detection would be consistent with the suggestion of \citet[][]{farnes17} that this feature is likely to be a low-level artefact from the deconvolution process.

Related to the previous paragraph, there are, however, some generally low-level artefacts in our LOFAR map, likely due to residual calibration and/or deconvolution errors. For example, $3.7$ arcmin to the north-north-west of SS\,433, at a position angle of $-20^{\circ}$, there is a spurious feature consisting of two components, with a peak flux density of approximately $80$ mJy beam$^{-1}$. A similar structure can be seen below the central shell of W 50 near the compact radio source NVSS~J191252+043104.

\subsection{The H\,{\sc ii} region S\,74}\label{nearby features} 

S\,74 \citep*[][]{crampton78}, a foreground H\,{\sc ii} region \citep[distance $\sim$$2$--$3$ kpc; e.g.][]{forbes89,brand93,lockman07,alves10} to the north-west of W\,50, is also detected at $140$ MHz. The peak flux density is $\sim$ $50 \pm 20$ mJy beam$^{-1}$, with the quoted $1\sigma$ error dominated by the uncertainty in the background level. In the convolved $1465$-MHz image from D98, as mentioned above in Section~\ref{W50 morphology}, we estimate a peak flux density of $75 \pm 10$ mJy beam$^{-1}$; thus, a representative two-point spectral index $\alpha^{1465}_{140}$ at the position of peak intensity is $0.2 \pm 0.2$. Although H\,{\sc ii} regions can become optically thick at low frequencies \citep*[e.g.][]{kassim89,copetti91,subrahmanyan95,omar02}, our estimated spectral index for S\,74 is consistent with the canonical spectral index of $-0.1$ in the optically-thin regime: the difference is only $1.5\sigma$. The radio spectrum may in fact be less inverted than our estimate depending on the relative contribution to the brightness temperature at $140$ MHz from the Galactic emission behind S\,74 (e.g. formulae in \citealt{nord06}; also see sky maps in e.g. \citealt{haslam82}, \citealt{guzman11}, \citealt{remazeilles15} and \citealt{zheng17}). 

In a $30.9$-MHz Galactic plane survey, conducted by \citet[][]{kassim88} with the Clark Lake Teepee-Tee Telescope, S\,74 appears as a negative absorption hole against the Galactic background (we refer the reader to \citealt[][]{nord06} and \citealt[][]{su17a,su17b} for other similar cases). There is evidence for such behaviour at $74$ MHz as well (J. Miller-Jones, priv. comm.). Although further measurements at low frequencies are needed to establish exactly where the spectral turnover occurs, we can make some predictions based on an estimated turnover frequency. Firstly, the linear emission measure $EM$ can be calculated approximately using the expression 
\begin{equation}\label{equation emission measure}
\tau_{\nu} \approx 3.28 \times 10^{-7} \left(\frac{T_{\rm e}}{10^4 \, \rm{K}}\right)^{-1.35} \left(\frac{\nu}{\rm{GHz}}\right)^{-2.1} \left(\frac{EM}{\rm{pc\,cm}^{-6}}\right),  
\end{equation} 
where $\tau_{\nu}$ is the optical depth at frequency $\nu$, and $T_{\rm e}$ is the electron temperature of the ionized gas in the H\,{\sc ii} region \citep[e.g.][]{mezger67}. Taking a turnover frequency of $\sim$$100$ MHz, where the optical depth is unity, and assuming $T_{\rm e} \sim 6000-7000$ K for S\,74 (e.g. see derived relationships between electron temperature and Galactocentric distance for H\,{\sc ii} regions in \citealt{shaver83}, \citealt*{paladini04}, and \citealt{alves12}), then $EM \sim 1.2$--$1.5 \times 10^4$ pc cm$^{-6}$. 

Secondly, due to the diffuse Galactic emission in the vicinity of the source, it is somewhat difficult to assess the full extent of S\,74 in our LOFAR map, but a rough estimate is $\sim$ $7$ arcmin $\times$ $13$ arcmin ($\sim$ $5.1$ pc $\times$ $9.5$ pc at a distance of $2.5$ kpc; also see e.g. \citealt{altenhoff70}, \citealt*{geldzahler80}, and \citealt{kuchar97} for other size estimates). Then, using the simple approximation that $EM \approx n_e^2 L$, where $n_{e}$ is the electron number density in S\,74, and $L$ is the path length through the source along the line of sight, $n_{\rm e}$ $\sim$ $40$--$45$ cm$^{-3}$ assuming that the path length is roughly the average of the projected dimensions on the sky (i.e. $L\sim7.3$ pc). Referring to the sub-classes in e.g. \citet[][]{kurtz05}, our very crude analysis suggests that S\,74 is most consistent with `classical' H\,{\sc ii} regions.

\section{Radio spectra}\label{spectral index discussion}

\begin{table*}
\begin{minipage}{180mm}
\centering
\caption{Flux densities and two-point spectral indices for the different components of W\,50. The $74$-MHz flux densities (J. Miller-Jones, priv. comm.) are on the same flux density scale as the LOFAR data (see Section~\ref{integrated flux densities spectra}); this scale is consistent with the one used in D98, to an accuracy within the typical uncertainties quoted in the table. The $327.5$-MHz flux densities were inferred using the $1465$-MHz flux densities given in D98, as well as the $\alpha^{1465}_{327.5}$ values, which were determined by D98 using a linear regression method. Flux density uncertainties at $327.5$ and $1465$ MHz have been estimated based on the uncertainties reported by D98 for the entire W\,50 nebula. At $327.5$ MHz, our total inferred flux density ($177 \pm 21$ Jy) is consistent with the value reported for the entire nebula in D98 ($160 \pm 20$ Jy). Apart from $\alpha^{1465}_{327.5}$, two-point spectral indices and the associated uncertainties were deduced from the integrated flux densities.}
  \begin{tabular}{cccccccccc}
  \hline
 Component & \multicolumn{1}{c}{$S_{74}$}  &  \multicolumn{1}{c}{$S_{140}$} & \multicolumn{1}{c}{$S_{327.5}$} & \multicolumn{1}{c}{$S_{1465}$} & $\alpha^{140}_{74}$ & $\alpha^{327.5}_{140}$ & $\alpha^{1465}_{327.5}$ & $\alpha^{1465}_{140}$  \\
 & \multicolumn{1}{c}{(Jy)} & \multicolumn{1}{c}{(Jy)} & \multicolumn{1}{c}{(Jy)} & \multicolumn{1}{c}{(Jy)} & &  & & \\   
  \hline
Eastern wing & \multicolumn{1}{r}{$61 \pm 5$} & \multicolumn{1}{r}{$48 \pm 13$} & \multicolumn{1}{r}{$51 \pm 6$} & \multicolumn{1}{r}{$15.0 \pm 1.5$} & \multicolumn{1}{r}{$-0.38 \pm 0.44$} & \multicolumn{1}{r}{$0.07 \pm 0.35$} & \multicolumn{1}{r}{$-0.82 \pm 0.05$} & \multicolumn{1}{r}{$-0.50 \pm 0.12$} \\
Central shell & \multicolumn{1}{r}{$245 \pm 16$} & \multicolumn{1}{r}{$121 \pm 25$} & \multicolumn{1}{r}{$99 \pm 12$} & \multicolumn{1}{r}{$43 \pm 4$} & \multicolumn{1}{r}{$-1.11 \pm 0.34$} & \multicolumn{1}{r}{$-0.24 \pm 0.28$} & \multicolumn{1}{r}{$-0.56 \pm 0.05$} & \multicolumn{1}{r}{$-0.44 \pm 0.10$}  \\
Western wing & \multicolumn{1}{r}{$54 \pm 4$} & \multicolumn{1}{r}{$38 \pm 8$} & \multicolumn{1}{r}{$27 \pm 3$} & \multicolumn{1}{r}{$13.0 \pm 1.3$} & \multicolumn{1}{r}{$-0.55 \pm 0.35$}  & \multicolumn{1}{r}{$-0.40 \pm 0.28$} & \multicolumn{1}{r}{$-0.49 \pm 0.09$} & \multicolumn{1}{r}{$-0.46 \pm 0.10$}   \\
\hline
\end{tabular}
\label{table:fluxes_components}
\end{minipage}
\end{table*}

\begin{figure*}
\begin{minipage}{180mm}
\centering
\includegraphics[width=18cm]{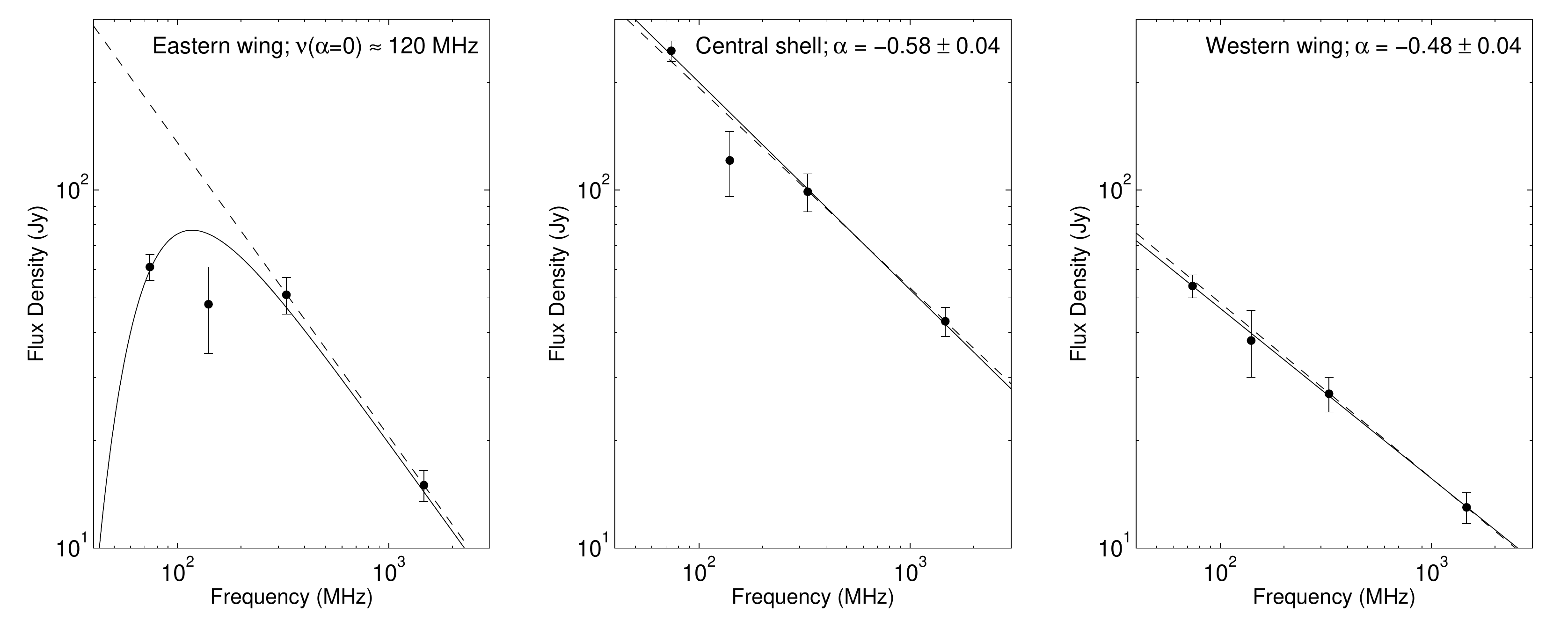}
\caption{Broadband radio spectra for the various components of W\,50, using the flux densities from Table~\ref{table:fluxes_components}. The solid lines are either a fit based on a free-free absorption model (left-hand panel; see Section~\ref{integrated flux densities spectra} for further details), or linear fits in $\log(S_{\nu})$--$\log(\nu)$ space (middle and right-hand panels). Inverse-variance weighting was used. Dashed lines represent radio spectra based on the two-point spectral indices $\alpha^{1465}_{327.5}$ from D98. For the eastern wing, we report the approximate turnover frequency, where the spectral index is zero, while we give the fitted spectral indices for the central shell and western wing. The full set of parameters can be found in Table~\ref{table:fluxes_components2}. Error bars are $\pm 1\sigma$.}
\label{fig:spectra}
\end{minipage}
\end{figure*}

As mentioned in Section~\ref{introduction}, \citet[][]{millerjones07} found that the integrated radio spectrum of W\,50 is well described by a single power law between $74$ and $4750$ MHz ($\alpha=-0.51 \pm 0.02$). Using their data at $74$ MHz, as well as the previously published data from D98 at $327.5$ and $1465$ MHz, they also demonstrated that the only component of W\,50 that shows evidence for significant spectral turnover at low frequencies is the eastern wing.

Our LOFAR $140$-MHz map not only bridges the frequency gap in the \citet[][]{millerjones07} analysis, but has sufficient dynamic range and angular resolution to permit us to study changes in the spectral index across the W\,50 nebula, either using the integrated flux densities for the specific components, or pixel by pixel. We now discuss our findings using a combination of both approaches.

\subsection{Integrated flux densities}\label{integrated flux densities spectra} 
 
Table~\ref{table:fluxes_components} lists the integrated flux densities for the eastern wing, central shell and western wing at $74$, $140$, $327.5$ and $1465$ MHz. While our LOFAR data are on the \citet[][]{roger73} flux density scale, the $74$-MHz observations used in the \citet[][]{millerjones07} study were calibrated with the \citet[][]{baars77} scale. Hence, we took the $74$-MHz flux densities for the three components (J. Miller-Jones, priv. comm.), scaling each measurement up by $10$ per cent and propagating the corresponding uncertainty term of $6$ per cent with the previous uncertainty; these values were found to be appropriate in an analogous rescaling of the $74$-MHz VLSSr \citep[][]{lane14}.      

In Table~\ref{table:fluxes_components}, we also give two-point spectral indices calculated from the integrated flux densities, although we note that the $\alpha^{1465}_{327.5}$ values are from D98 and were calculated with a linear regression method using the pixel values in specified regions of the source (further details in Section~\ref{spectral index discussion map}). In addition, broadband radio spectra are presented in Figure~\ref{fig:spectra}. 

As can be seen in Figure~\ref{fig:spectra}, our LOFAR flux densities are consistent with the previously published measurements, in particular strengthening the evidence for a spectral turnover in the eastern wing. We have fitted each spectrum with either a free-free absorption model (eastern wing), or a linear fit in $\log(S_{\nu})$--$\log(\nu)$ space (central shell and western wing). The best-fitting parameters are reported in Table~\ref{table:fluxes_components2}, and the fits are overlaid in the panels in Figure~\ref{fig:spectra}. The reduced $\chi^2$ value for the western wing ($0.035$) suggests that our uncertainties are possibly overestimated in this case. 

In the case of the eastern wing, our model is of the form    
\begin{equation}\label{eqn free free} 
S_{\nu} = S_{140,{\rm u}}\left(\frac{\nu}{140\,\rm{MHz}}\right)^{\alpha}\exp\left[-(\tau_{140})\left(\frac{\nu}{140\,\rm{MHz}}\right)^{-2.1}\right],
\end{equation} 
where $\tau_{140}$ is as before, $S_{140,{\rm u}}$ is the unabsorbed flux density at $140$ MHz for optically-thin emission, and $\alpha$ is the optically-thin spectral index \citep[e.g.][]{weiler86}. We assume that $\alpha=-0.82$, using the $\alpha^{1465}_{327.5}$ value from D98. The turnover frequency (at about $120$ MHz, where the fitted integrated flux density is approximately $80$ Jy) is near the bottom of the LOFAR high band, although more low-frequency measurements are needed to confirm this. Further discussion can be found in Section~\ref{alpha free-free}.    

\subsection{W\,50 spectral index map}\label{spectral index discussion map}

\begin{table}
\centering
  \caption{Values of the best-fitting parameters to the radio spectra in Figure~\ref{fig:spectra}. The units of $S$ and $\nu$ are Jy and MHz, respectively.}
  \begin{tabular}{lccc}
  \hline
\multicolumn{4}{l}{Free-free absorption fit:} \\  
\multicolumn{4}{l}{$S_{\nu} = S_{140,{\rm u}}\left(\frac{\nu}{140\,\rm{MHz}}\right)^{-0.82}\exp\left[-(\tau_{140})\left(\frac{\nu}{140\,\rm{MHz}}\right)^{-2.1}\right]$} \\  
\\
 & $S_{140,{\rm u}}$ (Jy) & $\tau_{140}$ & Reduced $\chi^2$ \\
Eastern wing & $99 \pm 8$ & $0.27 \pm 0.03$ & $2.57$  \\ 
\\\cmidrule{1-4}
\\
\multicolumn{4}{l}{Linear fit:}   \\
\multicolumn{4}{l}{$\log(S_{\nu}) = a + b\log(\nu)$}   \\
\\
 & $a$ & $b$ \, $(\equiv \alpha)$ & Reduced $\chi^2$ \\
Central shell & $3.46 \pm 0.09$ & $-0.58 \pm 0.04$ & $1.23$  \\ 
Western wing & $2.62 \pm 0.10$ & $-0.48 \pm 0.04$ & $0.035$  \\ 
\hline
\end{tabular}
\label{table:fluxes_components2}
\end{table}

\begin{figure*}
\begin{minipage}{180mm}
\centering
\includegraphics[width=17cm]{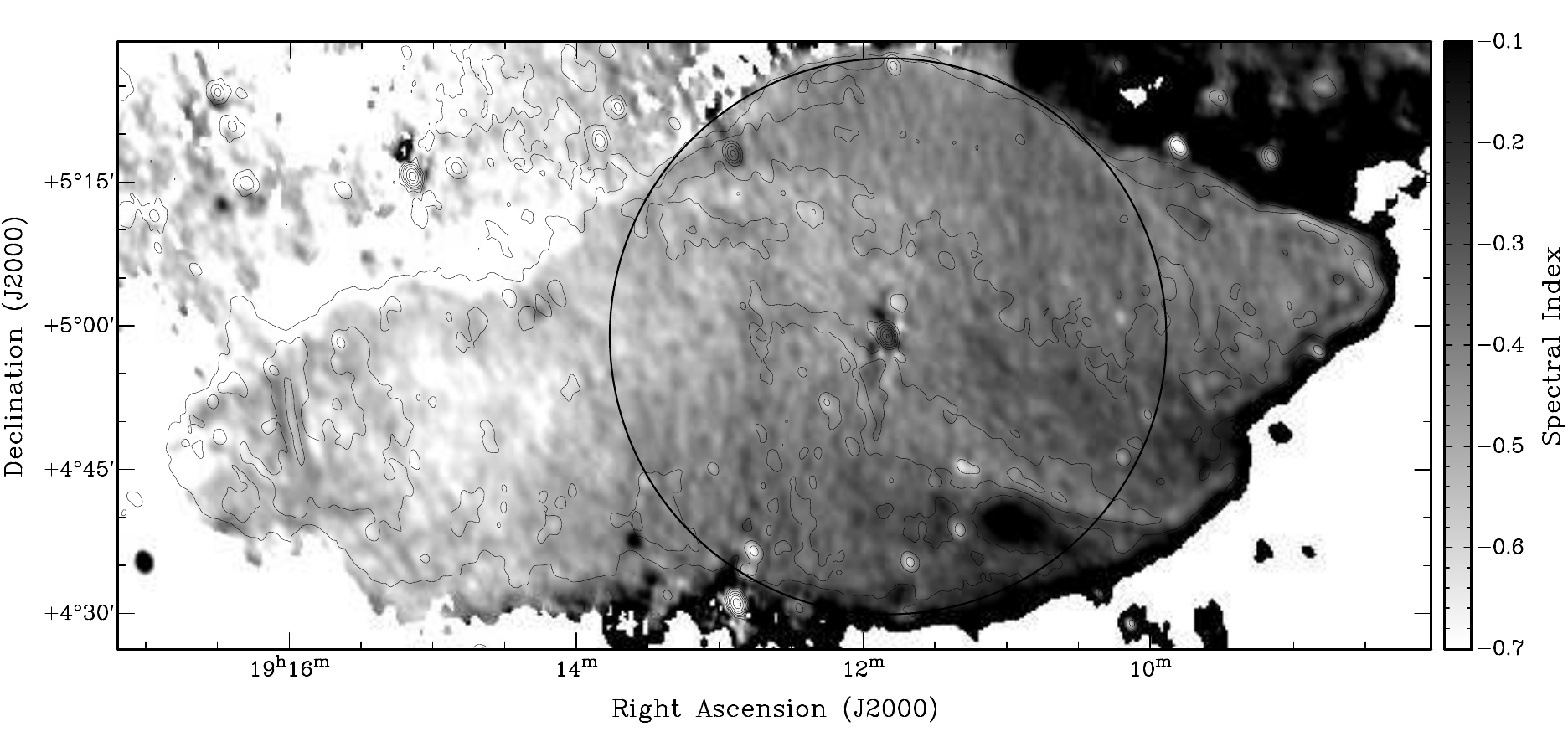}
\caption{Two-point spectral index map between $140$ and $1465$ MHz. Pixels with flux densities $S_{1465} < 5$ mJy beam$^{-1}$ have been masked out. LOFAR contours have also been overlaid to outline the shape of the nebula; the lowest contour is at $30$ mJy beam$^{-1}$ and each successive contour is a factor of two higher in surface brightness (up to a maximum of $1.92$ Jy beam$^{-1}$). In addition, as in Figure~\ref{fig:deepmap2}, we show the central shell. The pixel-by-pixel spectral indices across W\,50 range from about $-0.9$ to $0.2$, but, to ensure a suitable contrast range in the plot, we use a subset of this range (indicated in the greyscale bar). Average background levels of $-4.5$ mJy beam$^{-1}$ and $3$ mJy beam$^{-1}$ at $140$ and $1465$ MHz, respectively (see Section~\ref{W50 fluxes} and D98), were subtracted before the spectral index map was generated. However, note that the contours plotted are the raw values.}
\label{fig:alpha_map}
\end{minipage}
\end{figure*}
  
We generated a two-point spectral index map for W\,50 using the $140$-MHz image shown in Figures~\ref{fig:deepmap} and \ref{fig:deepmap2}, as well as the $1465$-MHz image from D98; see Figure~\ref{fig:alpha_map}. Firstly, to account for the fact that the angular resolutions are slightly different, we convolved the $1465$-MHz image (resolution $56$ arcsec $\times$ $54$ arcsec, BPA $-61.4^{\circ}$; figures 1a and 1b in D98), to the resolution of our LOFAR map ($78$ arcsec $\times$ $55$ arcsec, BPA $20^{\circ}$). Secondly, for consistency with the $1465$-MHz map, we regridded the LOFAR image from $10$ arcsec pixel$^{-1}$ to $15$ arcsec pixel$^{-1}$. Lastly, we calculated the spectral index values: $\log(S_{140}/S_{1465})/\log(140/1465)$ per pixel, with average background corrections of $-4.5$ and $3$ mJy beam$^{-1}$ at $140$ and $1465$ MHz, respectively, applied first. 

Generally, there is a very good agreement between the astrometry of the two maps, although there are some examples of non-systematic offsets of order $1$--$2$ pixels for moderately bright, compact sources in the field of view. Investigating this further, it appears as though our astrometry is generally more consistent with, for example, the $1.4$-GHz NVSS. Given the limitations in our analysis, which we describe below, we have not attempted to make differential alignment corrections to account for these relatively small offsets.

One difficulty with the method described above for calculating the spectral index map is that varying, non-zero background levels will cause biases; in particular, this is an issue for our LOFAR image. Hence, the results are best interpreted in a qualitative sense. As in the $\alpha^{1465}_{327.5}$ spectral index map (figure~4 in D98), we too observe regions of flatter spectral indices towards the south-west of the source, either side of the steeper, bright filament. However, in general, the spectral index gradients are not as pronounced in our map; moreover, while the apparent steepening from west to east across the source is consistent with the D98 spectral index map, this is, at least in part, likely to result from biases related to the background levels (see below). The brightest filamentary structure in the eastern wing may slightly flatten immediately to the east and west of the central region, although higher-resolution maps are needed to confirm this. The very flat spectral indices along some of the boundary regions are artificial: the negative residual backgrounds at $140$ MHz result in underestimated $140$-MHz flux densities. 

There is a hole-like feature approximately centred at RA $19^{\rm h}$\,$10^{\rm m}$\,$58\fs1$, Dec. $+04^{\circ}$\,$39\arcmin$\,$57\farcs8$ (J2000), in the south-western region of the central shell below the bright filament, that has an inverted spectrum (maximum value $\sim$0.2). In the LOFAR band, it is $\sim$$7$--$8$ times fainter than the diffuse emission surrounding it; the feature is labelled in Figure~\ref{fig:deepmap2}. In D98, figures 1b and 2 suggest possibly similar behaviour at $327.5$ and $1465$ MHz, but not as pronounced in the latter case (also see figure 1 in \citealt[][]{farnes17}). This may be a significantly absorbed region, although a corresponding foreground H\,{\sc ii} region at or near these coordinates could not be found in the SIMBAD \citep[][]{wenger00} and VizieR \citep*[][]{ochsenbein00} online databases.\footnote{http://simbad.u-strasbg.fr/simbad/ and http://vizier.u-strasbg.fr/viz-bin/VizieR} Moreover, \citet[][]{farnes17} presented a continuum-corrected H$\alpha$ mosaic of the W\,50 region, using data from the Isaac Newton Telescope Photometric H$\alpha$ Survey of the Northern Galactic Plane \citep[IPHAS;][]{drew05}; however, an excess of emission at the position of the feature is not observed (figure 6 in \citealt[][]{farnes17}).    

As mentioned above, generating a spectral index map using the standard formula, pixel by pixel, can potentially give highly misleading quantitative results. Therefore, following \citet[][]{anderson93}, as well as D98 (also see e.g. \citealt[][]{leahy91} and \citealt[][]{leahy98}), we investigated a linear regression method. Various box sizes ($5$--$35$ pixels, i.e. $75$ arcsec -- $8.75$ arcmin, per side) were slid over both maps in increments of one pixel in both directions. We then used the average slope from the regressions of $S_{1465}$ versus $S_{140}$ and $S_{140}$ versus $S_{1465}$ (taking the inverse of the calculated slope in the latter case) to determine the spectral index for each area, which was assigned to the central pixel. Regions off-source were masked out. Rather than considering the ratio of the flux densities at $140$ and $1465$ MHz, this approach, also often referred to as the $T$--$T$ plot method (where $T$ is the brightness temperature), removes any bias which is constant over the box size, but averages over the spectra of small-scale variations within the box (e.g. structure within filaments). Larger box sizes will have a range of faint and bright emission to provide more accurate constraints when determining each slope. 

After using the technique described in the previous paragraph, the apparent $\alpha^{1465}_{140}$ spectral index gradient visible in Figure~\ref{fig:alpha_map} is significantly reduced. Each of the three components of W\,50 has about the same mean spectral index, $\sim$$-0.4$, a value consistent with the two-point spectral indices $\alpha^{1465}_{140}$ derived from the integrated flux densities in Table~\ref{table:fluxes_components}. This mean is not affected significantly by the choice of the dimensions of the box, apart from the typical $1\sigma$ error, which decreases from $\sim$$0.3$ to $\sim$$0.1$ with increasing box size. The effects of correlated noise, which can, for example, be reduced by only considering pixels at regularly spaced intervals \citep[e.g.][]{green90}, are second order here. Unlike between $327.5$ and $1465$ MHz, an approximately constant average spectral index $\alpha^{1465}_{140}$ across W\,50 simply follows from the fact that there is spectral turnover in the eastern wing at low frequencies (Table~\ref{table:fluxes_components} and Figure~\ref{fig:spectra}), and so $\alpha^{1465}_{140}$ is significantly flatter than $\alpha^{1465}_{327.5}$ for the eastern wing.    

\subsection{Spectral index asymmetry between the wings}\label{spectral index asymmetry} 

D98 suggested that the density gradient due to the nearby Galactic plane can explain the difference in the radio spectra of the two wings between $327.5$ and $1465$ MHz: the western wing has a flatter spectrum because the jet is expanding into a denser region, leading to stronger shocks and higher levels of compression. However, our results in Section~\ref{spectral index discussion map} are less consistent with this interpretation, albeit with the possible complication of absorption effects at low frequency. On the other hand, in Figure~\ref{fig:alpha_map}, the steepening of the spectral index from west to east across W\,50 is a relatively smooth. Some of this is highly likely to be due to the corresponding gradient in the background level at $140$ MHz, but we cannot determine with certainty whether there is a residual gradient after accounting for the varying background levels: the uncertainties in the spectral index maps generated with the linear regression method are too large.

We now explore an alternative explanation for any spectral index asymmetry between the wings. Expanding into the higher-density environment closer to the Galactic plane, the western wing has been less affected by adiabatic expansion, and so its magnetic field will consequently be stronger than that in the more expanded eastern wing of the nebula. Electrons emitting synchrotron radiation at a specific frequency $\nu$ in an ambient magnetic field of strength $B$ have a Lorentz factor $\gamma$ given by 
\begin{equation}\label{lorentz factor} 
\gamma = \left( \frac{2 \pi m_{\rm e} \nu}{e B} \right)^{1/2},
\end{equation}
where $e$ is the elementary charge and $m_{\rm e}$ is the electron rest mass. Therefore, in the western wing, where the magnetic field is stronger, we would probe lower Lorentz factors at a fixed radio frequency. Moreover, for a curved distribution of the underlying electron Lorentz factors, we should also measure a flatter spectral index \cite[e.g.][]{scheuer68}.   

If this hypothesis is correct, then we may expect the equipartition magnetic field $B_{\rm eq}$ to be stronger in the western wing. However, X-ray observations of W\,50 suggest similar ranges of $B_{\rm eq}$ in the eastern and western wings: \citet[][]{safi97} calculated values of $7$--$60$ $\upmu$G (eastern wing) and $7$--$49$ $\upmu$G (western wing; also see \citealt[][]{moldowan05}). The equipartition magnetic fields can also be estimated from radio data, although the calculation can involve significant uncertainties; revisions to the classical approach \citep[e.g.][]{longair94} have been suggested \citep[e.g.][]{beck05,arbutina12,arbutina13,duffy12}. In particular, \citet[][]{arbutina12,arbutina13} developed modified equipartition calculations for the particular case of SNRs, although it is possible that further modifications would be needed for W\,50, given both its distinct morphology and the relatively flat spectral index of the western wing (their model is valid for $-1 \leq \alpha \leq -0.5)$.    

While it is beyond the scope of this paper to carry out a detailed comparison between the different methods, we now briefly consider the formalism developed by \citet[][]{duffy12}. Despite the fact that the focus of their study is the lobes of radio galaxies and quasars, the framework described is based upon fundamental synchrotron physics. The key quantities in equation (26) of \citet[][]{duffy12} are as follows: (i) a spectral curvature term (that is intrinsic curvature and not from absorption effects); (ii) the frequency of the peak of the radio spectrum; and (iii) the emissivity at this frequency. All terms vary slowly, however, due to power-law dependencies of $1/7$ or $2/7$. We assume that the majority of the radio emission in the eastern wing can be modelled as a cylinder, height $47$ arcmin and diameter $22$ arcmin; similarly, we model the western wing as a cone with height $25$ arcmin and base diameter $27$ arcmin. Some straightforward calculations can be made to show that, very tentatively, and in a qualitative sense, $B_{\rm eq}$ may be at least slightly stronger in the western wing if the associated spectral curvature is milder than for the eastern wing. However, the comparison will be affected by spectral curvature that is caused by foreground free-free absorption (Section~\ref{alpha free-free}).

In addition, despite the limitations mentioned above, given that the radio spectrum of the western wing could have a spectral index slightly steeper than $-0.5$ within the $1\sigma$ uncertainties (Table~\ref{table:fluxes_components2}), then the online calculator\footnote{http://poincare.matf.bg.ac.rs/{\raise.17ex\hbox{$\scriptstyle\sim$}}arbo/eqp/} for the \citet[][]{arbutina12,arbutina13} model also possibly suggests a slightly stronger value of $B_{\rm eq}$ for the western wing, if the non-spherical geometries of both wings are compensated for, to first order. Potentially significant differences in the depths and filling factors of the two wings are additional sources of uncertainty, both for this model and the \citet[][]{duffy12} formalism. More complete radio spectra are also needed.

\subsection{Low-frequency spectral curvature}\label{alpha free-free} 

Below 327.5 MHz, it is the eastern wing and not the western wing that has a flatter spectrum; this is contrary to our suggestion in Section~\ref{spectral index asymmetry}. One possibility is that the underlying electron energy spectra in the two wings are different, with perhaps shock acceleration in the denser ambient environment of the western wing accounting for the lack of spectral curvature, at least over the frequency range considered in this paper. Alternatively, the radio spectrum of the eastern wing may correspond to enhanced low-frequency free-free absorption along the line of sight to this component of the nebula. This hypothesis is supported by the fact that \citet[][]{farnes17} found a global anti-correlation for W\,50 between the linearly polarized flux density at $2.3$ GHz and the strength of the diffuse H$\alpha$ emission, suggesting that this is due to a foreground Faraday screen of warm, ionized plasma (also see the polarization measurements in \citealt*[][]{downes81b}, and \citealt*[][]{downes86}). Inspecting the H$\alpha$ mosaic from this study (their figure 6), bright emission is coincident with the position of the eastern wing. 

In Section~\ref{integrated flux densities spectra}, we fitted the radio spectrum of the eastern wing with a free-free absorption model. Taking the fitted value $\tau_{140} = 0.27 \pm 0.03$ (Table~\ref{table:fluxes_components2}), and using Equation~\ref{equation emission measure} with the assumption that $T_{\rm e} \sim 10^4$K, then the corresponding linear emission measure is $EM \sim 1.3 \times 10^4$ pc cm$^{-6}$. However, as mentioned before, more low-frequency data points are needed (e.g. within the LOFAR high band), particularly in light of the relatively poor reduced $\chi^2$ value of $2.57$.  

The potential effects of a synchrotron ageing break also need to be considered. Radiative losses from the emitting electrons result in a break in the synchrotron spectrum, and, for continuous particle injection, the spectrum is steepened from $\nu^{\alpha}$ to $\nu^{\alpha-0.5}$. The break frequency $\nu_{\rm B}$ is given by
\begin{equation}
\nu_{\rm B} = \frac{1610^2\,{\rm GHz}}{(B/\upmu{\rm G})^3(t_{\rm s}/{\rm Myr})^2},
\end{equation}
where $t_{\rm s}$ is the `synchrotron age' of the distribution, i.e. the time since it was a power law out to infinite frequency  \citep[e.g.][]{carilli91}. If the age of W\,50 is approximately $20\,000$ yr \citep[][]{goodall11a}, then a break frequency of e.g. $\sim$$100$--$300$ MHz would require a magnetic field of strength $\sim$$2800$--$4000$ $\upmu$G. These values are $1.7$--$2.8$ dex larger than the equipartition magnetic fields that were discussed above in Section~\ref{spectral index asymmetry}. Therefore, assuming that the lobes do not deviate significantly from equipartition, the effects of synchrotron ageing can be neglected. 

In the $30.9$-MHz Galactic plane survey by \citet[][]{kassim88}, two thirds of a sample of $32$ SNRs exhibit low-frequency spectral turnovers \citep[][]{kassim89a,kassim89b}. W\,50 was detected, although it is not part of the study presented in \citet[][]{kassim89a,kassim89b}; the eastern half of the source falls outside of the survey area, due to the Galactic latitude cutoff imposed. The brightest part of the western wing and the northern part of the central shell are clearly detected, with corresponding flux densities of $35.2$ and $23.7$ Jy, respectively (flux density uncertainty approximately $20$ per cent; angular resolution $13.0$ arcmin $\times$ $11.1$ arcmin). However, the more southerly regions of the nebula are at best marginally detected. Given that the $30.9$-MHz flux densities are on the \citet[][]{baars77} scale, we adjusted the $74$-MHz flux densities in Table~\ref{table:fluxes_components} back to this scale (dividing by a factor of $1.1$; see Section~\ref{integrated flux densities spectra}), as well as making an additional correction for the $74$-MHz flux density of the central shell: about $25$ per cent of the flux density orginates from the same spatial region as at $30.9$ MHz (J. Miller-Jones, priv. comm.). We therefore obtain approximate spectral indices $\alpha^{74}_{30.9}$ for the western wing and central shell of $\sim$$0.4$ and $\sim$$1.0$, respectively. These spectral indices, coupled with the distinct change in morphology, provide further evidence for a low-frequency spectral turnover in W\,50, possibly due to free-free absorption along the line of sight. LOFAR low-band measurements, which cover the frequency range $30$--$80$ MHz, would allow a far more detailed investigation to be made.  

Finally, we briefly consider whether a spectral turnover at low frequencies could also be related to a low-frequency cutoff in the electron energy spectrum. In this case we would expect a $S_{\nu} \propto \nu^{0.3}$ dependence \citep[e.g.][]{scheuer68}. Using Equation~\ref{lorentz factor} and assuming the equipartition magnetic field strengths calculated by \citet[][]{safi97}, $\gamma \sim 420$--$1240$ at $30$ MHz. It is not entirely clear whether a low-frequency cutoff would therefore be discernible with LOFAR, particularly given the upper end of this range \citep[e.g. discussion in][]{harris05}. Further measurements between $30$ and $80$ MHz would also potentially allow contributions from free-free absorption, a low-frequency cutoff and intrinsic spectral curvature to be disentangled.   

\section{Low-frequency variability of SS\,433}\label{SS433 variability} 

\subsection{150-MHz light curve}\label{SS433 variability_low} 

\begin{figure*}
\begin{minipage}{180mm}
\centering
\includegraphics[width=12cm]{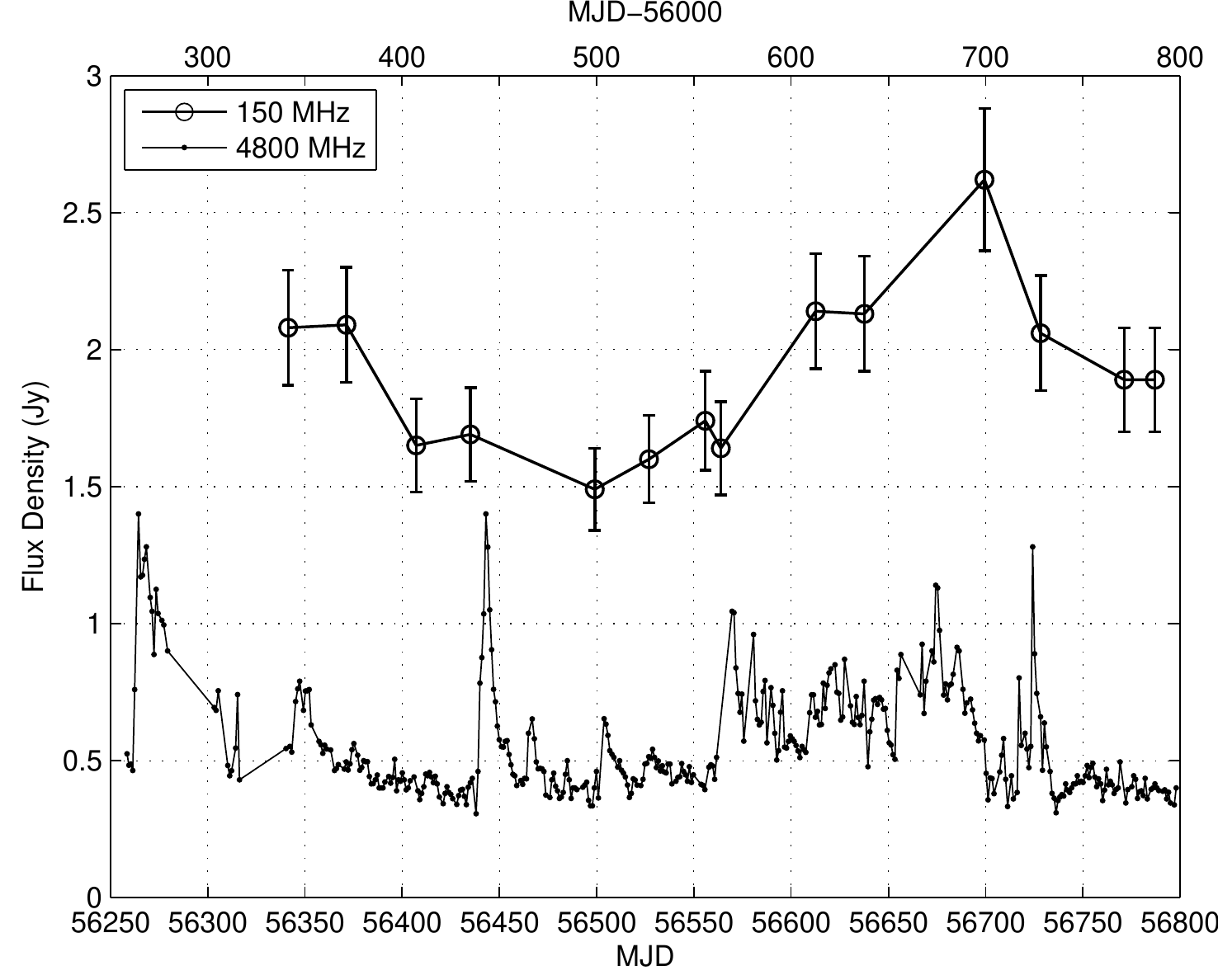}
\caption{LOFAR $150$-MHz light curve for SS\,433 over the period 2013 February -- 2014 May, as well as the RATAN-600 $4800$-MHz light curve from 2012 November -- 2014 May. We show $\pm1\sigma$ error bars for each LOFAR data point; these are dominated by the $\pm 10$ per cent calibration uncertainty. The uncertainty for each $4800$-MHz data point is $\pm5$ per cent.}
\label{fig:lightcurve}
\end{minipage}
\end{figure*}

A light curve of SS\,433 from all $14$ observations (Table~\ref{table:observations}) is shown in Figure~\ref{fig:lightcurve}. For the deep run, we have not used the SS\,433 flux density reported in Section~\ref{deep map}. Instead, we made another averaged map from the same baseline range as that used when imaging the monitoring runs, and over a similar, more restricted $uv$ coverage. We then divided by a correction factor of $1.2$ to shift the effective average frequency from $140$ to $150$ MHz, bringing the flux density scale into line with the monitoring observations (see Section~\ref{in band flux scale} for further details). 

For the monitoring runs and the regenerated map from the first run described above, the angular resolution is about $140$ arcsec $\times$ $100$ arcsec (median BPA $\approx$ $60^{\circ}$). Each Gaussian fit to the emission from SS\,433 was point-like. Note that the new measured flux density from the deep run (as given in Table~\ref{table:observations}) agrees with the value given in Section~\ref{deep map} within the $1\sigma$ uncertainties.     

In Figure~\ref{fig:lightcurve}, SS\,433 shows tentative evidence for variability in the LOFAR band. To quantify its magnitude and significance, we used two statistics. Firstly, the modulation index $m$ is given by 
\begin{equation}
m = \frac{\sigma_{S}}{\overline{S}},
\end{equation} 
where $\sigma_{S}$ is the standard deviation of the combined set of flux densities from Table~\ref{table:observations}, and $\overline{S}$ is the mean flux density. We find that $m$ is $16$ per cent, which is comparable to the modulation index at $235$ MHz \citep[$21$ per cent;][]{pandey07}. If we take into account the errors on each data point and debias the modulation index (as, for example, is described in \citealt{bell14}), then $m$ is reduced to $11$ per cent. 

Secondly, we calculated the $\chi^2$ probability that the flux density remained constant over the period covered by our observations \citep[e.g.][]{gaensler00,bell14}. The quantity 
\begin{equation}\label{eqn_light_curve_constant_check}
x^2 = \sum\limits_{i=1}^n \frac{(S_{i} - \tilde{S})^2}{\sigma^2_i},
\end{equation} 
where $S_{i}$ is the $i$th flux density measurement, $\tilde{S}$ is the weighted mean flux density, and $\sigma_i$ is the uncertainty for the $i$th flux density measurement, follows a $\chi^2$ distribution with $n-1$ degrees of freedom, assuming that the uncertainties are drawn from a normal distribution. The weighted mean is calculated using the formula
\begin{equation}\label{eqn_light_curve_constant_check2}
\tilde{S} = \sum\limits_{i=1}^n \left(\frac{S_{i}}{\sigma^2_i}\right) \bigg/ \sum\limits_{i=1}^n \left(\frac{1}{\sigma^2_i}\right).
\end{equation} 
The assumed 10 per cent calibration uncertainty is the dominant error term. Background corrections off-source, including from W\,50, are second-order effects because of the $uv$ range chosen, and hence we do not consider the small contributions from the associated uncertainties.     

Using Equations~\ref{eqn_light_curve_constant_check} and \ref{eqn_light_curve_constant_check2} for the $n=14$ observations in Table~\ref{table:observations}, we find that the probability $P$ of $x^2$ exceeding our calculated value by chance (i.e. $P(x^2 > 28.3)$ for $13$ degrees of freedom) is $8.2 \times 10^{-3}$. This probability value may be too large to be regarded as a significant indicator of variability \citep[e.g. the criteria discussed in][]{gaensler00}, but we suggest that it is sufficiently small that it could be interpreted as being at least marginally significant. Furthermore, in Figure~\ref{fig:lightcurve}, there are hints of a general consistency in the calibration for several of the more closely spaced observations, suggesting that the calculated $\chi^2$ probability does not simply result from a light curve with large scatter, but rather genuine variability. Reducing the calibration uncertainty further would be needed to better quantitatively assess the magnitude of the intrinsic variability of the source.

\subsection{Comparison with 4800-MHz data}\label{SS433 variability_high} 

In Figure~\ref{fig:lightcurve}, we have also plotted $4800$-MHz flux densities from contemporaneous monitoring observations (angular resolution $5$ arcmin $\times$ $50$ arcsec) carried out with the RATAN-600 telescope.\footnote{http://www.sao.ru/cats/cgi-bin/ss433.cgi} The same broad trends are observed at both $150$ and $4800$ MHz: an initial decrease in flux density to a relatively constant baseline level (approximately $1.6$ and $0.44$ Jy at $150$ and $4800$ MHz, respectively), followed by increased activity at later epochs, before a second decline in flux density at the end of the sampling period. 

\subsubsection{Individual flares}\label{individual flares}

At $4800$ MHz, there were bright flares on MJD $56442.99$ and $56724.22$, which peaked at $1.40$ and $1.28$ Jy, respectively. However, neither of them have obvious counterparts at $150$ MHz. There was also activity before and shortly after our LOFAR observing campaign began, with a maximum $4800$-MHz flux density of $1.40$ Jy on MJD $56264.48$. 

We now discuss whether we would have expected to see the outbursts in the LOFAR band. Firstly, if the plasmons associated with a flare are well described by, for example, the \citet[][]{vanderlaan66} synchrotron bubble model (also see e.g. \citealt{hjellming88}, and \citealt{ball93}), then the rise time of an outburst is longer at lower frequencies (owing to opacity effects), and the peak flux density is often lower (relative to the quiescent baseline level). The ratio of the peak flux densities at $150$ and $4800$ MHz is given by
\begin{equation}\label{equation vdL}
\frac{S_{{\rm max},150}}{S_{{\rm max},4800}} =  \left(\frac{4800}{150}\right)^{-(7p+3)/(4p+6)},
\end{equation}
where $p = 1-2\alpha$ is the negative of the power-law index of the electron energy ($E$) distribution (i.e. $N(E)\mathrm{d}E \propto E^{-p}\mathrm{d}E$). We use a fiducial value of $1$ Jy for the $4800$-MHz flares (after subtracting the quiescent component), and, for simplicity, assume that the spectral index is in the range $-1$ and $0$, which is broadly consistent with long-term monitoring data between $960$~MHz and $21.7$~GHz from RATAN-600 \citep[][]{trushkin03}. Then, for the resulting values of $p$ ($1$--$3$), we would expect the peak flux density at $150$ MHz to be about $1$--$3$ per cent of the $4800$-MHz value, i.e. approximately $10$--$30$ mJy. This is much less than the typical calibration uncertainty (Table~\ref{table:observations}). Conversely, a hypothetical flare of, say, $500$ mJy at $150$ MHz, marginally detectable given our current calibration uncertainty, would be $\sim$tens of Jy at $4800$ MHz. Such a bright outburst has not been observed before from SS\,433. 

An alternative approach is as follows. Firstly, \citet[][]{trushkin03} empirically derived an average power-law relationship between the maximum flux density $S_{\rm max}$ of a flare (after the quiescent component has been subtracted) and the observing frequency $\nu$, such that 
\begin{equation}\label{equation flare flux}
S_{\rm max} = 1.3 \nu^{-0.46},
\end{equation}
where $S_{\rm max}$ is in units of Jy, and $\nu$ is in GHz. Equation~\ref{equation flare flux} is inconsistent with the predictions of the \citet{vanderlaan66} model for the values of $p$ that we considered above; this equation only agrees with Equation~\ref{equation vdL} for a significantly inverted spectral index $\alpha \approx 0.83$. However, a negative power-law dependence is, for example, expected in the model developed by \citet*[][]{marti92}, at frequencies where the outburst is initially optically thin; this model assumes particle injection into twin jets that are adiabatically expanding in the lateral direction, exponentially at first. 

From Equation~\ref{equation flare flux}, a hypothetical $1$-Jy flare at $4800$ MHz is brighter than what we would expect on average at this frequency ($0.63$ Jy). Nonetheless, assuming that the average power-law index still holds, a $1$-Jy flare at $4800$ MHz would be about $2$ Jy at $960$ MHz, the lowest RATAN-600 frequency used in \citet[][]{trushkin03}. Moreover, in the absence of a spectral turnover, an extrapolation to lower frequencies would suggest a flux density of around $5$ Jy in the LOFAR high band. Despite our limited sampling, there is no evidence of such a large flux density increase at $150$ MHz in Figure~\ref{fig:lightcurve}.       

However, if we assume that a flare from SS\,433 will initially be optically thick below $\sim$$1$ GHz, then we can use equations (29) and (30) in \citet[][]{marti92} to estimate the flux density ratio of the flare between $150$ and $960$ MHz. These equations are
\begin{equation}\label{eqn_marti_ff}
S_{\rm max} \propto \nu^{(2.78p+3.46)/9}
\end{equation}
in the case of free-free absorption, and 
\begin{equation}\label{eqn_marti_ssa}
S_{\rm max} \propto \nu^{(13p+2)/(7p+8)}
\end{equation}
for synchrotron self-absorption. For the value of $p$ corresponding to Equation~\ref{equation flare flux}, i.e. $p=1.92$, we would therefore expect the $\sim$$2$-Jy flare at $960$~MHz to have a peak flux density of either $\sim$$340$ or $\sim$$200$~mJy at $150$~MHz. Repeating the entire alternative approach for $p = 1$--$3$, including modifying the spectral index from Equation~\ref{equation flare flux} accordingly, the expected peak flux density at $150$ MHz varies between $\sim$$160$--$440$ mJy for a $1$-Jy flare at $4800$ MHz. The flare, although now brighter than was predicted using Equation~\ref{equation vdL}, would still not be detectable with the current calibration uncertainty.        

Also note that as a rough consistency check, if one were to start with the average maximum flux density at $960$ MHz from Equation~\ref{equation flare flux}, i.e. $1.3$ Jy, and do the same analysis as described in the previous paragraph, then the expected peak flux density at $408$ MHz would be either $\sim$$560$ or $\sim$$440$~mJy. On average, flares at $408$ MHz have an excess flux density of $1110 \pm 410$ mJy above the quiescent baseline level \citep[][]{bonsignori86}, which overlaps with our calculated values within at most $1.6\sigma$ (also see \citealt[][]{vermeulen93}). If the initial transition between optically-thin and optically-thick flares occurs below $\sim$$1$ GHz, then the flares at $408$ MHz would generally be brighter than predicted above using Equations~\ref{eqn_marti_ff} and \ref{eqn_marti_ssa}, and the chances of a detection at $150$ MHz would also increase. 

The fact that the $150$- and $4800$-MHz observations were independent observing programmes also needs to be considered. \citet[][]{trushkin03} empirically derived a power-law relationship between the time at which the maximum flux density occurs during a flare, $t_{S_{\rm max}}$, and the observing frequency, such that 
\begin{equation}\label{equation flare delay}
t_{S_{\rm max}} = 5.1 \nu^{-0.43},
\end{equation}
where $t_{S_{\rm max}}$ is in units of days, and $\nu$ is in GHz. Equation~\ref{equation flare delay} suggests a delay of about $9$ days between $4800$ and $150$ MHz, assuming that it is still valid at low frequencies. The first LOFAR observation after the $4800$-MHz flare on MJD $56442.99$ occurred $56$ days afterwards, while there was a LOFAR observation $4.1$ days after the flare on MJD $56724.22$. While in theory we may have been able to detect the flare rise with LOFAR in the latter case, the extended $4800$-MHz activity before this particular flare may complicate the analysis at low frequencies (Section~\ref{extended flaring}). Closely coordinating observations across a wide range of radio frequencies will be the most robust approach for studying future flaring activity from SS\,433. 

\subsubsection{Extended flaring activity}\label{extended flaring} 

The onset of $4800$-MHz activity at MJD $\sim$56560 is accompanied by a sustained $\sim$$0.5$--$1$ Jy rise in flux in the LOFAR band. There is also a hint of a similar flux density increase when comparing the $150$- and $4800$-MHz data taken at times up to and including the epoch corresponding to the second LOFAR data point. It is possible that the numerous separate high-frequency bursts have become blended together at $150$ MHz, owing to the longer associated rise time-scales at low frequencies. This is also very tentatively suggested by the apparent difference between the $150$-MHz flux densities at the start and end of the $4800$-MHz flaring activity beginning at MJD $\sim$56560: approximately $1.63$ and $1.89$ Jy, respectively. 

The median peak brightness of these flares at $4800$ MHz is about $360$ mJy, after correcting for the quiescent levels on either side of the period of activity. A few tens of flares, spaced $\sim$5 days apart, would not replicate the observed rise in the LOFAR flux density if the \citet[][]{vanderlaan66} model is assumed: a plateau can be modelled, but it is over an order of magnitude too faint and decays too rapidly. Although beyond the scope of this paper, if individual flares are instead much brighter at $150$ MHz, as, for example, was suggested above in Section~\ref{individual flares} when considering the \citet[][]{marti92} model, then a sustained rise of order several hundred mJy or more is a possibility. More complex scenarios previously outlined in the literature \citep*[e.g.][]{vermeulen93,blundell11,jeffrey16} may also be relevant. 

\subsubsection{Possible effects of refractive interstellar scintillation}\label{scintillation} 

Another potential explanation for the relatively slow variability that we observe at $150$ MHz is refractive interstellar scintillation \citep[RISS; e.g.][]{rickett86}. \citet[][]{pandey07} investigated whether RISS could explain the variability of SS\,433 in their GMRT data at $235$ and $610$ MHz, concluding that the source is most likely affected by this phenomenon. Taking their calculated time-scales for RISS and making a standard assumption that this time-scale is proportional to $\lambda^{2.2}$ (where Kolmogorov turbulence is assumed), then the associated time-scale at $150$ MHz is $\sim$$2.3$ yr. This value is roughly double the length of our monitoring campaign, and so it seems unlikely that we are seeing a corresponding effect in our data. A much longer time baseline at $150$ MHz would be beneficial.     
    
\subsection{Quiescent flux densities} 

Considering the data points between MJD $56407$--$56560$, before the onset of the extended flaring activity, the median $4800$-MHz flux density, $0.44$ Jy, is close to the long-term average quiescent value of $0.43$ Jy (\citealt{trushkin03}; also note that the median flux density at the end of the flaring activity, $0.40$ Jy, is very similar). In addition, the average flux density of the five LOFAR observations over this period is $1.63 \pm 0.07$ Jy, which is consistent with both $160$ MHz quiescent flux density measurements from \citet[][]{seaquist80}: $1.6 \pm 0.2$ Jy (1979 July 1) and $2.2 \pm 0.3$ Jy (1979 July 21 and 22). These data were obtained at a similar angular resolution to the LOFAR observations ($2.3$ arcmin $\times$ $1.9$ arcmin; BPA $0^{\circ}$) and also have an absolute flux density scale that is consistent with our data \citep[][]{roger73,seaquist80,scaife12}. 
 
\citet[][]{seaquist80} suggested that the quiescent radio spectrum of SS\,433 turns over near $300$ MHz, although there was a gap in their data between $160$--$408$ MHz. In Table~\ref{table:quiescent_fluxes}, we list the average flux densities from this study, along with more recent measurements, including our LOFAR data. We have estimated the quiescent flux densities at $235$ and $610$ MHz from the monitoring data presented by \citet[][]{pandey07}; given the possibility of flaring activity and/or the effects of interstellar scintillation during this observing campaign (discussion in \citealt[][]{pandey07}), our estimate in each case is the range between the minimum flux density and the median flux density. 

\begin{figure}
\centering
\includegraphics[width=8.55cm]{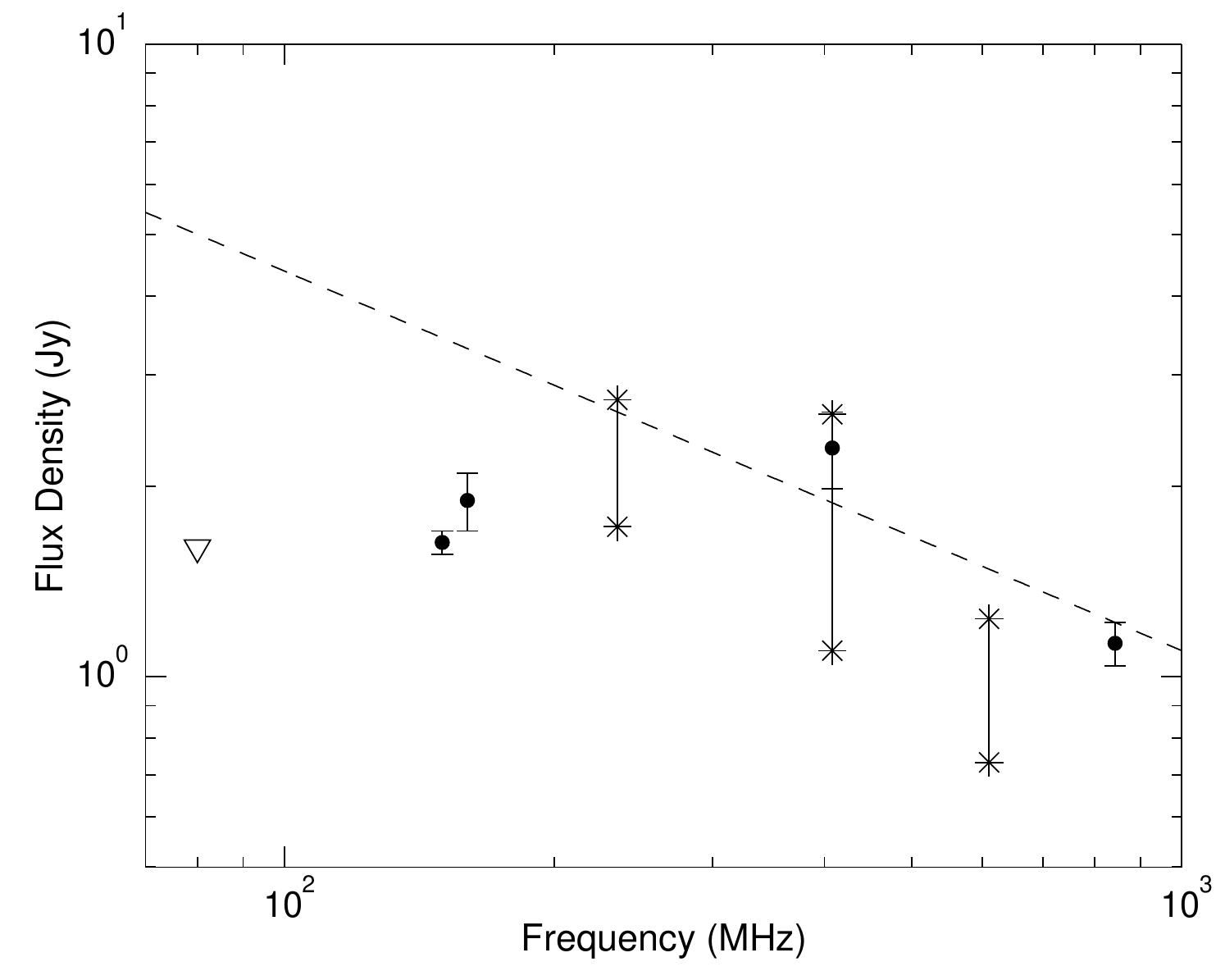}
\caption{The low- and mid-frequency radio spectrum for SS\,433 during quiescence, using the flux densities from Table~\ref{table:quiescent_fluxes}. The dashed line is the RATAN-600 quiescent flux density empirical power-law fit from \citet[][]{trushkin03}, i.e. $S_{\nu} = 1.1\nu^{-0.6}$ (where $S_{\nu}$ and $\nu$ are in units of Jy and GHz, respectively), extrapolated to lower frequencies. Error bars indicate $\pm 1\sigma$ uncertainties, the solid lines between asterisks indicate flux density ranges at $235$, $408$ and $610$ MHz, and the triangle represents the average $5\sigma$ upper limit at $80$ MHz.}
\label{figure:quiescent_fluxes}
\end{figure}

Figure~\ref{figure:quiescent_fluxes} shows the quiescent radio spectrum at low and mid frequencies. We have also plotted an extrapolated empirical power-law fit that was derived by \citet[][]{trushkin03} using higher-frequency RATAN-600 data, where the spectral index $\alpha = -0.60 \pm 0.14$. The spectrum still appears to be consistent with a turnover at around $300$ MHz, as suggested by \citet[][]{seaquist80}, and our LOFAR data further suggest that the spectrum has truly turned over at $150$--$160$ MHz. However, there may be discrepancies resulting from the different angular resolutions of the various datasets, which would affect the relative level of contamination from W\,50. Again, in-band LOFAR flux densities could be used to obtain a more complete picture of the low-frequency properties of the quiescent radio spectrum of SS\,433. It may also be particularly useful to obtain new HBA measurements at the upper end of the operating frequency range ($210$--$240$ MHz) and GMRT data at $325$ MHz.

\begin{table}
 \centering
  \caption{Quiescent flux density estimates for SS\,433 at low and mid frequencies. We have calculated average values at $80$ and $160$ MHz based on the individual measurements presented in \citet[][]{seaquist80}. In addition, the ranges we have calculated at $235$ and $610$ MHz are based on the minimum and median flux densities in \citet[][]{pandey07}. The $843$-MHz value was determined from the tabulated flux densities in \citet[][]{vermeulen93}. Note that the upper limit at $80$ MHz is approximately consistent with the $74$-MHz flux density from \citet[][]{millerjones07}, although there are not enough higher-frequency data available to fully establish the extent of activity from SS\,433 when the $74$-MHz data were obtained (J. Miller-Jones, priv. comm.).}
  \begin{tabular}{ccc}
  \hline
   \multicolumn{1}{c}{Frequency}  &  \multicolumn{1}{c}{$S_{\rm{SS}433}$} & \multicolumn{1}{c}{Reference(s)} \\    
\multicolumn{1}{c}{(MHz)} & \multicolumn{1}{c}{(Jy)}  \\
 \hline
 $80$ & $<1.6^{a}$ & (1) \\
 $150$ & $1.63 \pm 0.07$ & (2) \\
 $160$ & $1.9 \pm 0.2$ & (1) \\
 $235$ & $1.73$--$2.74$ & (3) \\
 $408$ & $1.1$--$2.6$ & (4,1) \\
 $408$ & $2.30 \pm 0.32$ & (5) \\
 $610$ & $0.73$--$1.23$ & (3) \\
 $843$ & $1.13 \pm 0.09$ & (6) \\ 
\hline
\multicolumn{3}{l}{References: (1) \citet[][]{seaquist80}; (2) this paper;} \\ 
\multicolumn{3}{l}{(3) \citet[][]{pandey07}; (4) \citet[][]{spencer79};} \\ 
\multicolumn{3}{l}{(5) \citet[][]{bonsignori86};} \\
\multicolumn{3}{l}{(6) \citet[][]{vermeulen93}.} \\
\multicolumn{3}{l}{$^{a}$$5\sigma$ upper limit.} \\
\end{tabular}
\label{table:quiescent_fluxes}
\end{table}

\subsection{Variability of nearby sources}\label{variability nearby sources} 

We searched for transients, as well as other variable sources in the field, by running all of our LOFAR maps through the {\sc transients pipeline} \citep[{\sc trap};][]{swinbank15}. No transients were found over the course of our monitoring campaign. Moreover, SS\,433 has the most significant variability statistics of all the detected sources. Given the discussion in Section~\ref{SS433 variability_low}, our data are therefore not sufficiently precise to allow us to draw conclusions concerning the potential low-frequency variability of other sources in the field.    

D98 detected a $\sim$$340$-mJy point source at $327.5$ MHz, close to the bright filaments in the eastern wing of W\,50 (coordinates RA $19^{\rm h}$\,$15^{\rm m}$\,$49.8^{\rm s}$, Dec. $+04^{\circ}$\,$51\arcmin$\,$45\arcsec$; J2000). We have labelled the position of this source with a white cross in Figure~\ref{fig:deepmap2}. Given several non-detections or marginal detections at higher frequencies, D98 suggested that it may be variable. We do not detect this source in any of our LOFAR datasets (e.g. $3\sigma$ upper limit $\sim$$25$ mJy beam$^{-1}$ in the deep observation, after taking into account the background level from W\,50). Furthermore, the source appears not to be present in the \citet[][]{farnes17} ATCA map. There are also non-detections in four $327.5$-MHz VLA maps over the period 2001 January--October, as well as the 1996 August 19 run used in the D98 study; the $3\sigma$ upper limit from the map generated using all five observations is $\sim$$17$ mJy beam$^{-1}$  (J. Miller-Jones, priv. comm.). This result suggests that the source may have only been present in the second $327.5$-MHz dataset that D98 obtained on 1997 August 18. Another possibility may be that the source is an artefact generated during the imaging process described in D98, which involved source subtraction and re-addition steps; see section 2.2 in D98.

\section{Candidate supernova remnant G\,38.7$-$1.4}\label{SNR candidate}  

\begin{figure*}
\begin{minipage}{180mm}
\centering
\includegraphics[width=9.0cm]{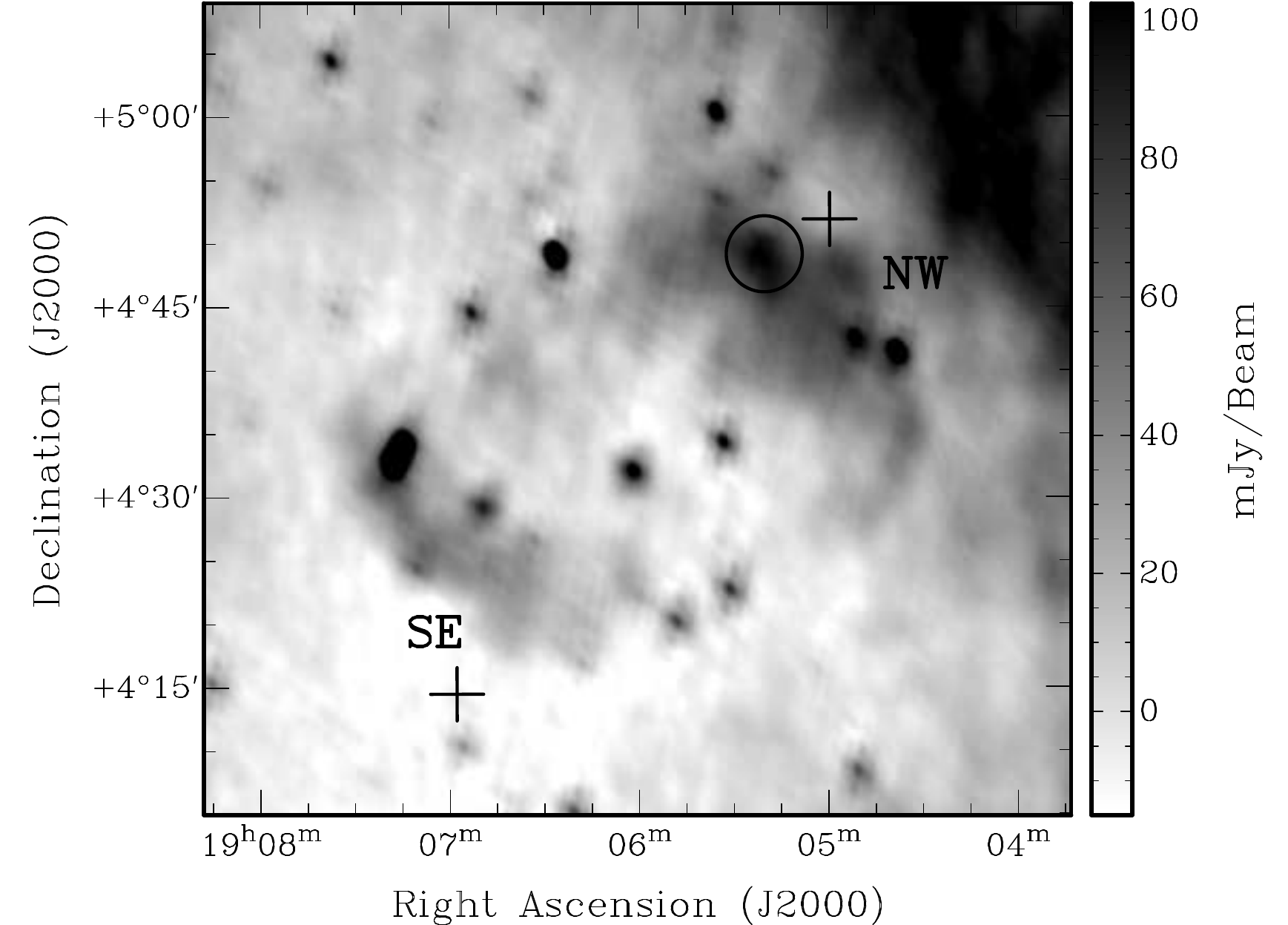}
\hspace{1cm}
\includegraphics[width=7.8cm]{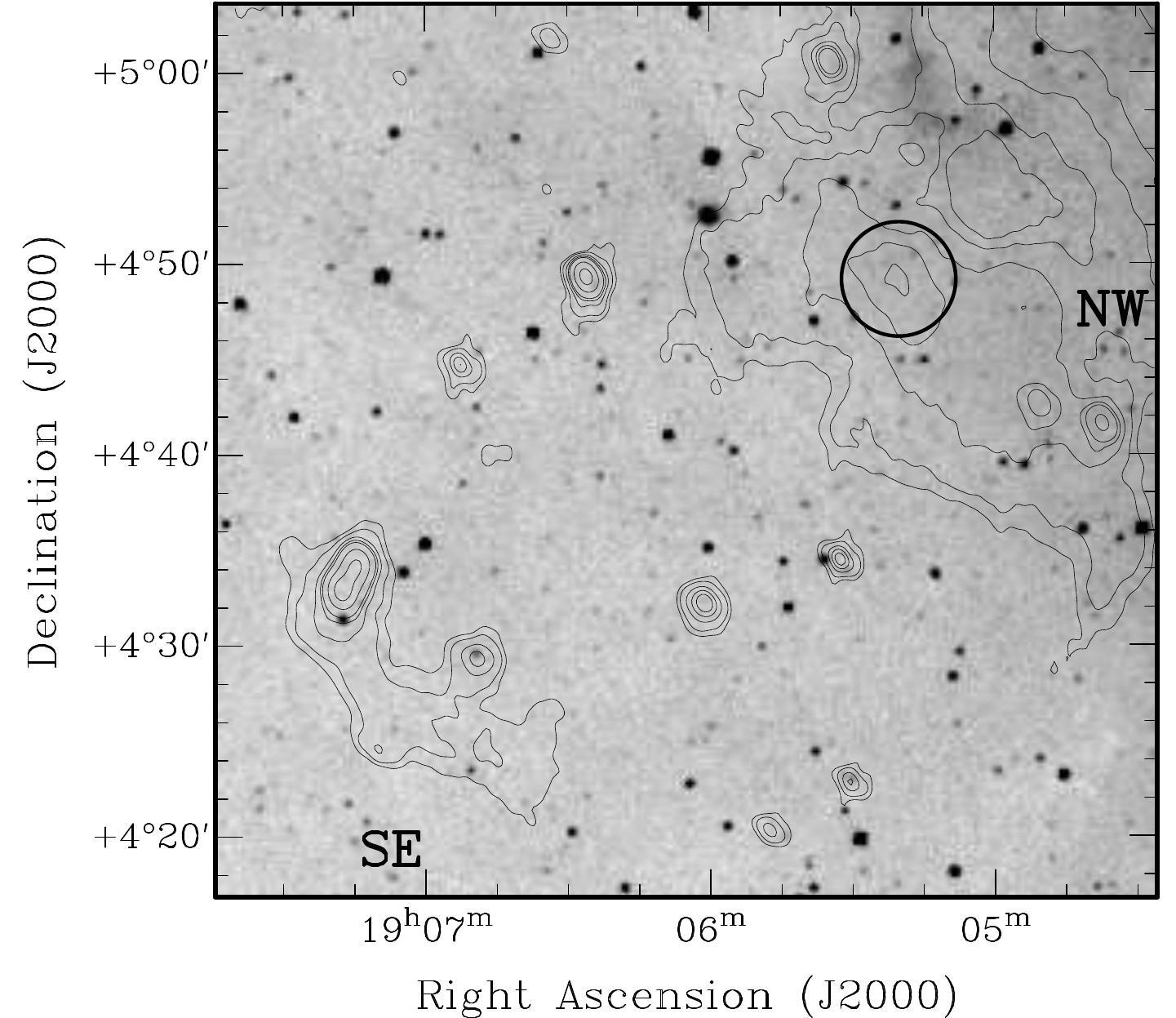}
\caption{{\em Left}: LOFAR detection of the radio shell of G\,38.7$-$1.4. The north-western (NW) and south-eastern (SE) components of the shell are labelled to assist the discussion in the text, and the knot from which a spectral index was estimated is circled. The positions of PSR~J1904$+$0451 and PSR~J1906$+$0414 are indicated by crosses. An average background level correction of $-40$ mJy beam$^{-1}$ has been applied to the map. {\em Right}: MSX $8$-$\upmu$m image of the G\,38.7$-$1.4 region, overlaid with LOFAR contours indicating surface-brightness levels of $30$, $40$, $60$, $80$, $100$, $200$ and $400$ mJy beam$^{-1}$ (also background corrected). As in the left-hand panel, the knot from which a spectral index was estimated is circled. There is no infrared counterpart to the nebula. Note that compared to the left-hand panel, the right-hand panel shows a slightly more zoomed-in view on G\,38.7$-$1.4.}
\label{fig:snr}
\end{minipage}
\end{figure*}

In Figure~\ref{fig:deepmap}, west-south-west of SS\,433 and W\,50, there is a diffuse radio source that is coincident with the candidate SNR G\,38.7$-$1.4 (also known as SNR G\,38.7$-$1.3). This candidate was first suggested by \citet{schaudel02}, who searched the \textit{ROSAT} all-sky survey \citep{voges99} for extended X-ray sources that could be unidentified SNRs. They found a $12$ arcmin $\times$ $8$ arcmin source that is coincident with an incomplete radio shell observed in the Effelsberg Galactic Plane Survey at $11$ cm \citep{reich84,reich90}, favouring a SNR interpretation. An independent identification was also made at optical wavelengths in the IPHAS H$\alpha$ survey by \citet{sabin13}, who further supported their claims through comparisons with the same radio shell, as seen in other archival radio datasets. 

Additional evidence for a SNR identification was found by \citet{huang14}. Using \textit{XMM-Newton} and \textit{Chandra X-ray Observatory} observations, these authors suggested that the X-ray emission is from shock-heated plasma. In addition, their analysis of archival VLA radio observations at $1.4$ GHz revealed a coincident broken radio shell, consistent with that reported by \citet{schaudel02}. \citet[][]{huang14} estimated that G\,38.7$-$1.4 is at a distance of $\sim$$4$ kpc, with an age of $\sim$$14\,000$ yr, and also suggested that the source belongs to the SNR mixed-morphology category \citep{rho98}. 

Although previously detected as a broken radio shell, our deep observation of SS\,433 and W\,50 has sufficient surface-brightness sensitivity and angular resolution to reveal a more complete radio shell morphology (left-hand panel of Figure~\ref{fig:snr}). In particular, the candidate supernova remnant is significantly larger than had been indicated by the X-ray observations, with a diameter of approximately $40$ arcmin $\times$ $30$ arcmin, centred at Galactic coordinates $(l,b) = (38\fdg7, -1\fdg1)$. However, accurate flux density measurements are hampered by the fact that the source sits in a significant negative bowl. After making a first-order background correction and assuming that the compact sources coincident with the shell are unrelated, the average $140$-MHz surface brightness of the diffuse emission is $\sim$$30$ mJy beam$^{-1}$. 

Traditionally, the confirmation of a SNR is achieved by demonstrating that the radio emission is non-thermal and/or linearly polarized. A polarimetric analysis of our LOFAR dataset was not possible at the time of writing. Also, while the south-eastern part of the shell (labelled `SE' in the left-hand panel of Figure~\ref{fig:snr}) falls within the ATCA $2.3$-GHz mosaic from the \citet{farnes17} study, no associated linearly polarized emission is detected, which is suggested to be due to foreground depolarization. Therefore, the potential non-thermal nature of G\,38.7$-$1.4 can instead be tested by using the selection criteria outlined in \citet{brogan06}: other than having a shell-like or partial-shell-like morphology, the candidate must have a negative spectral index (where $S_{\nu} \propto \nu^{\alpha}$, as before) and must show a lack of associated mid-infrared $8$-$\upmu$m emission. 

The VLA Galactic Plane Survey \citep[VGPS;][]{stil06}, conducted at $1420$ MHz with $60$-arcsec angular resolution, covers the region containing G\,38.7$-$1.4. While only a hint of the south-eastern part of the shell is detected in the VGPS map, the north-western part of the shell (labelled `NW' in the left-hand panel of Figure~\ref{fig:snr}) is slightly brighter. Therefore, in order to calculate a rough spectral index for G\,38.7$-$1.4, we first measured the flux density from the brightest knot of diffuse radio emission that we assume to be associated with the north-western part of the shell, indicated in the left-hand panel of Figure~\ref{fig:snr}. Our estimate is $\alpha^{1420}_{140} \sim -0.8 \pm 0.1$, where the $1\sigma$ uncertainty is dominated by the uncertainty in the background subtraction in the LOFAR map. The fainter diffuse emission in the vicinity of this feature has a spectral index ranging from about $-1.0$ to $-0.5$, although the $1\sigma$ spectral index uncertainty can be as large as about $0.3$. Moreover, using a $3 \sigma$ upper limit at $1420$ MHz, an approximate spectral index constraint for the south-eastern part of the shell is $\alpha^{1420}_{140} \lesssim -0.6$. Therefore, given these various values, there is no significant evidence to suggest a discrepancy in relation to the general SNR spectral index distribution (e.g. $\overline{\alpha} \approx -0.48$ and $\sigma_{\alpha} \approx 0.15$  in the \citealt[]{green14} SNR catalogue), and, in addition, our measurements are indicative of non-thermal radiation. 

Referring to the third selection criterion from \citet[][]{brogan06}, at the position of G\,38.7$-$1.4, there is no evidence for associated mid-infrared emission at $8$ $\upmu$m, using Galactic plane survey data collected by the \textit{Midcourse Space Experiment} \citep[\textit{MSX};][]{price01}. The MSX map is shown in the right-hand panel of Figure~\ref{fig:snr}. This result further suggests a non-thermal nature for the shell, strengthening the potential SNR identification.

We also investigated the possibility of nearby pulsar associations by searching the Australia Telescope National Facility Pulsar Catalogue \citep{manchester05}.\footnote{http://www.atnf.csiro.au/people/pulsar/psrcat/} The two closest known pulsars are PSR~J1904$+$0451 (angular distance $\approx$ 21 arcmin from the centre of the nebula), with a $1400$-MHz flux density of $0.117$ mJy, and PSR~J1906$+$0414 (angular distance $\approx$ $26$ arcmin), with a $1400$-MHz flux density of $0.23$ mJy. Their positions are indicated in the left-hand panel of Figure~\ref{fig:snr}. At $140$ MHz, the $3\sigma$ flux density upper limits are $\sim$$15$--$20$ mJy beam$^{-1}$ (background corrected). PSR~J1904$+$0451 is an isolated millisecond pulsar with a spin period of $6.09$ ms; the dispersion measure distance and characteristic age are $4.7$ kpc (using the \citealt{cordes02} model) and $1.7 \times 10^{10}$ yr, respectively \citep[][]{scholz15a,scholz15b}. PSR~J1906$+$0414 is also not part of a binary system and has a spin period of $1.04$ s; furthermore, the dispersion measure distance and characteristic age are $7.5$ kpc and $1.4 \times 10^{6}$ yr, respectively \citep[][]{lorimer06}. While there is certainly a very large discrepancy between the characteristic ages of the two pulsars and the estimated age of the SNR candidate, the potentially significant uncertainties associated with the various age and distance estimates do not allow any strong conclusions concerning a pulsar-SNR association to be made \citep[e.g. following the guidelines in][]{kaspi96}. Moreover, assuming an age of roughly $14\,000$ yr, an association in either case would imply that the pulsar has a projected velocity in the plane of the sky of up to several thousand km~s$^{-1}$, well into the tail of the velocity distribution for isolated pulsars \citep*[e.g.][]{arzoumanian02}. 

\citet{brogan06} demonstrated through $330$-MHz VLA observations of the Galactic plane that deep, mid-frequency surveys can significantly increase the number of known SNRs, helping to solve the long-standing problem of `missing' SNRs in the Galaxy \citep[e.g.][]{helfand89,brogan04}. Our detection of G\,38.7$-$1.4 with LOFAR is an encouraging result in this sense: although this source was previously known, we have shown that low-frequency observations with competitive low-surface-brightness sensitivity can also aid in the detection of diffuse emission. This is particularly so given that the surface brightness values will generally increase with decreasing frequency for the SNR spectral index distribution stated above. Galactic plane surveys conducted with the current generation of low-frequency radio telescopes therefore have the potential to reveal a hitherto missing population.

\section{Conclusions}\label{conclusions}

In this paper, we have presented the results from a LOFAR high-band observing campaign of the X-ray binary SS\,433 and its associated supernova remnant W\,50. We have drawn the following conclusions from our study:
\begin{enumerate}
\item Our $140$-MHz wide-field image of SS\,433 and W\,50 is the deepest low-frequency map of this system to date. The morphology of W\,50, as well as the flux densities of the various components, are consistent with previous investigations. Foreground free-free absorption may explain the curvature of the radio spectrum of the eastern wing, although more low-frequency data points are needed to confirm this result.  
\item We have tentatively detected variability from SS\,433, with both a debiased modulation index of $11$ per cent and a $\chi^2$ probability of a flat light curve of $8.2 \times 10^{-3}$. The $\sim$$0.5$--$1$ Jy rise in the $150$-MHz flux density may correspond to extended flaring activity over a period of about six months, observed at $4800$ MHz in contemporaneous RATAN-600 observations. The flux density increase is too large to explain with a standard synchrotron bubble model, although a model along the lines of the one presented in \citet[][]{marti92}, involving laterally expanding twin jets, may potentially offer an alternative explanation. While LOFAR could be a key trigger for other facilities if there is a prolonged period of activity, low-frequency detections of individual flares will require more accurate calibration, as well as higher-cadence sampling.   
\item Within the large field of view, a significant amount of extended structure has been detected near and along the Galactic plane. We have described the most complete detection thus far of the radio shell of SNR candidate G\,38.7$-$1.4. 
\end{enumerate}

Our findings demonstrate the potential of LOFAR, as well as other low-frequency Square Kilometre Array precursors and pathfinders, for studying not only X-ray binaries, but supernova remnants as well.

\section*{Acknowledgements}

We thank the anonymous referee for their insightful, detailed comments, which were valuable in refining our analysis and presentation. We also acknowledge the considerable efforts of the ASTRON Radio Observatory, particularly Carmen Toribio, in setting up the observations and pre-processing the data. In addition, we warmly thank George Heald and Christian Kaiser for useful discussions, as well as Gloria Dubner for kindly providing us with a $1465$-MHz map of SS\,433 and W\,50.  

JWB, RPF, AJS, GEA, TDS and MP acknowledge support from European Research Council Advanced Grant 267697 `4 Pi Sky: Extreme Astrophysics with Revolutionary Radio Telescopes'. AR, JDS, AJvdH, DC, GJM and RAMJW acknowledge support from European Research Council Advanced Grant 247295 `AARTFAAC'. JCAM-J is the recipient of an Australian Research Council Future Fellowship (FT140101082). SAT is grateful to the Russian Base Research Foundation for financial support (grant number 12-02-0812a) and acknowledges support through the Russian Government Program of Competitive Growth of Kazan Federal University. RPB acknowledges support from the European Research Council under the European Union’s Horizon 2020 research and innovation programme (grant agreement no. 715051; `Spiders'). SC acknowledges financial support from the UnivEarthS Labex program of Sorbonne Paris Cit\'e (ANR-10-LABX-0023 and ANR-11-IDEX-0005-02). JWTH acknowledges funding from an NWO Vidi fellowship and European Research Council Starting Grant ``DRAGNET'' (337062). JvL acknowledges funding from the European Research Council under the European Union's Seventh Framework Programme (FP/2007-2013) / ERC Grant Agreement n. 617199. The financial assistance of the South African SKA Project (SKA SA) towards this research is hereby acknowledged. Opinions expressed and conclusions arrived at are those of the authors and are not necessarily to be attributed to the SKA SA.

This paper is based (in part) on data obtained with the International LOFAR Telescope (ILT). LOFAR \citep[][]{vanhaarlem13} is the Low Frequency Array designed and constructed by ASTRON. It has facilities in several countries, that are owned by various parties (each with their own funding sources), and that are collectively operated by the ILT foundation under a joint scientific policy.  

This research has made use of the SIMBAD database and the VizieR catalogue access tool, operated at CDS, Strasbourg, France. This research has also made use of the NASA/IPAC Extragalactic Database (NED) which is operated by the Jet Propulsion Laboratory, California Institute of Technology, under contract with the National Aeronautics and Space Administration. 

The National Radio Astronomy Observatory is a facility of the National Science Foundation operated under cooperative agreement by Associated Universities, Inc. 

This research made use of data products from the \textit{Midcourse Space Experiment}. Processing of the data was funded by the Ballistic Missile Defense Organization with additional support from NASA Office of Space Science. This research has also made use of the NASA/ IPAC Infrared Science Archive, which is operated by the Jet Propulsion Laboratory, California Institute of Technology, under contract with the National Aeronautics and Space Administration.

\bsp

\label{lastpage}

\end{document}